\documentclass[pra,showpacs,eqsecnum]{revtex4}%
\usepackage{amsfonts}
\usepackage{amsmath}
\usepackage{bm}
\usepackage{amssymb}
\usepackage{graphicx}%
\begin{document}
\preprint{ }
\title[Negative index materials]{Macroscopic Maxwell's equations and negative index materials}
\author{B. Gralak}
\affiliation{Institut Fresnel, CNRS, Aux-Marseille Universit\'{e}, Ecole Centrale
Marseille, Campus de St J\'{e}r\^{o}me, 13397 Marseille Cedex 20, France}
\author{A. Tip}
\affiliation{FOM-Instituut AMOLF, Science Park 104, 1098 XG Amsterdam, The Netherlands}

\begin{abstract}
We study the linear phenomenological Maxwell's equations in the presence of a
polarizable and magnetizable medium (magnetodielectric). For a dispersive,
non-absorptive, medium with equal electric and magnetic permeabilities
$\varepsilon(\omega)$ and $\mu(\omega)$, the latter can assume the value $-1$
($+1$ is their vacuum value) for a discrete set of frequencies $\pm\hat
{\omega}_{n}$, i.e., for these frequencies the medium behaves as a negative
index material (NIM). We show that such systems have a well-defined time
evolution. In particular the fields remain square integrable (and the
electromagnetic energy finite) if this is the case at some initial time. Next
we turn to the Green's function $\mathsf{G}(\mathbf{x,y},z)$ (a tensor),
associated with the electric Helmholtz operator, for a set of parallel layers
filled with a material. We express it in terms of the well-known scalar $s$
and $p$ ones. For a half space filled with the material and with a single
dispersive Lorentz form for $\varepsilon(\omega)=\mu(\omega)$ we obtain an
explicit form for \textsf{G}. We find the usual behavior for negative index
materials for $\omega=\pm\hat{\omega}$, there is no refection outside the
evanescent regime and the transmission (refraction) shows the usual NIM
behavior. We find that \textsf{G} has poles in $\pm\hat{\omega}$, which lead
to a modulation of the radiative decay probability of an excited atom. The
formalism is free from ambiguities in the sign of the refractive index.

\end{abstract}
\date{\today}
\keywords{Maxwell's equations, NIM}
\pacs{03.50.De,78.20.Ci, 42.25.Gy}
\maketitle

\section{ Introduction}

Often magnetization plays a minor role in situations where the
phenomenological Maxwell's equations apply. But in recent years negative index
materials (NIM's), also called left handed materials, have become of
increasing interest, in particular due to the work of Veselago
\cite{Veselago1} and Pendry \cite{Pendry}. Here the magnetization is not
negligible at all. In general a NIM system is defined by the property that for
certain frequencies $\omega$ the electric permeability (permittivity)
$\varepsilon(\omega)$ or the magnetic permeability $\mu(\omega)$ becomes
negative. Of particular interest is the case where both become negative at the
same frequency $\hat{\omega}$, the NIM frequency, and are equal to the
opposite of their vacuum value, i.e., $-1$ instead of $+1$. Below we refer to
this case as the NIM situation.

For an introduction, containing an extensive set of references, see
\cite{Veselago2} (there are also NIM systems based upon specific properties of
photonic crystals, which are not considered here).

The existence of NIM's has been debated in the theoretical literature at
various occasions \cite{Existence}. In particular the sign of the index of
refraction, which involves taking a square root, has been a subject of
discussion. Naively it equals $+1$, in both vacuum and a NIM system but this
result is challenged for the NIM situation. For experimental verification, see
\cite{Experimental}. Calculations based upon a simple model, where one part of
space is vacuum ($\varepsilon=\mu=1$) and the other filled with a NIM
($\varepsilon=\mu=-1$, frequency-independent) , tend to give ambiguous
results. This is sometimes remedied by adding a small imaginary part to one of
the permeabilities but on the whole the situation is rather unclear.

The use of the phenomenological Maxwell's equations should solve possible
ambiguities but it seems that so far this approach has not been taken and here
we intend to fill this gap. Since fabricated materials, intended to study NIM
behavior, are usually anisotropic, we take the (space and frequency dependent)
susceptibilities (which relate the polarization and magnetization to the
electric and magnetic fields) to be tensors rather than scalars.

The first matter to be solved is the existence of a proper time evolution. In
view of the time convolutions in the constitutive equations this is not
directly obvious. The next task is to see if a NIM situation can exist. This
being the case, the following point of interest is obtaining the Helmholtz
Green's function and scattering amplitudes for specific configurations. The
former is important since its imaginary part enters the radiative decay rate
of an atom or nanostructure close to the material. Hence experimental results
on such decay rates can give information about the properties of the material.
In addition the Green's function, or rather the associated transition
operator, is required to describe scattering phenomena, such as reflection and
transmission in layered systems.

Thus we start with the phenomenological Maxwell's equations with general
frequency-dependent permeability tensors, satisfying the usual causality and
passivity conditions. After providing some relevant background and a summary
of the properties of the electric and magnetic susceptibilities we introduce
the auxiliary field formalism (AFF). The latter was presented earlier by one
of us (AT in \cite{LAD}) for dielectrics ($\mu=1)$. The idea is to introduce
an additional set of fields, the auxiliary fields, to remove the time
convolutions in Maxwell's equations. This has a number of advantages:

\noindent1) The combined set of electromagnetic and auxiliary fields satisfies
a unitary time evolution, thus insuring a proper time evolution for the
electromagnetic fields.\newline2) Such a system is easily quantized, leading
to a second quantization formalism that is rigorously valid for both
absorptive and dispersive systems.\newline3) The formalism implies that the
inverses of the electric and magnetic Helmholtz operators exist as bounded
operators, so the associated Green's functions are square integrable.\newline%
4) Setting up a scattering formalism is straightforward.

The AFF leads to a proper time evolution, notwithstanding the possibility that
for specific frequencies $\hat{\omega}$ we can have a NIM situation,
$\varepsilon(\hat{\omega})=\mu(\hat{\omega})=-1$. In case the initial fields
are square integrable they remain so for all later times. In the Appendix we
give a rigorous proof of this important fundamental property.

Another relevant piece of information is that the susceptibilities for general
dispersive, non-absorptive, systems consist of a (possibly infinite) sum of
Lorentz contributions (AT in \cite{Drude-Lorentz}). This immediately gives a
positive answer to the existence of NIM's. In case we are dealing with a
single dispersive Lorentz contribution%
\begin{equation}
\varepsilon(\omega)=\mu(\omega)=1-\frac{\Omega^{2}}{\omega^{2}-\omega_{0}^{2}%
},\label{1.1}%
\end{equation}
we note that for the NIM frequences $\omega=\pm\hat{\omega}$, $\hat{\omega
}^{2}=\omega_{0}^{2}+\Omega^{2}/2$ we have $\varepsilon(\pm\hat{\omega}%
)=\mu(\pm\hat{\omega})=-1$. Thus, theoretically, the NIM case can be realized
for dispersive, non-absorptive, systems, contrary to what is sometimes claimed
\cite{Existence}. Adding more Lorentz terms gives more frequency values with
this property but it remains a discrete set and in between the values of
$\varepsilon(\omega)$ and $\mu(\omega)$ vary wildly. In fact, between two
subsequent NIM frequencies there is always a frequency for which total
reflection takes place (for $\omega=\omega_{0}$ in the above example),
precisely the opposite of the NIM case, where no reflection is thought to be
the situation. Thus it seems that obtaining an extended frequency interval for
which the permeabilities are approximately equal to $-1$ is not possible. In
fact it was already noted by Veselago \cite{Veselago1} that a system showing
NIM behavior must be dispersive.\textbf{ }In case there is absorption,
$\varepsilon(\omega)=\mu(\omega)=-1$ cannot be realized for real $\omega$ as
can be seen by adding absorption to the above case,%

\begin{equation}
\varepsilon(\omega)=\mu(\omega)=1-\frac{\Omega^{2}}{\omega^{2}+i\gamma
\omega-\omega_{0}^{2}}. \label{1.2}%
\end{equation}
Next we introduce the Laplace-transformed Maxwell's equations and the tensor
Green's function $\mathsf{G}(\mathbf{x,y},z)$ related to the electric
Helmholtz operator. $\mathsf{G}(\mathbf{x,y},z)$ features most of the
properties of the system. As said the radiative decay rate of excited atoms is
proportional to its imaginary part. We then turn to layered systems and
express $\mathsf{G}(\mathbf{x,y},z)$ into a set of two scalar ones, for $s$
and $p$ polarization, respectively. Subsequently the Green's function for the
half space case, mentioned above, is studied. In particular we obtain an
explicit expression for $\mathsf{G}(\mathbf{x,y},z)$. Then $%
{\bm{E}}%
(\mathbf{x},t)$ is given by the inverse Laplace transform of%
\begin{equation}%
{\hat{\bm{E}}}%
(\mathbf{x,}z)=\int d\mathbf{y}\mathsf{G}(\mathbf{x,y},z)\cdot\mathbf{g(y},z),
\label{1.3}%
\end{equation}
where $\mathbf{g(y},z)$ is some square integrable initial field configuration
or an external current density. At this point the square root of
$\ z^{2}\varepsilon(\mathbf{x},z)\mu(\mathbf{x},z)-\kappa^{2}$, with $%
{\bm{\kappa}}%
$ a two-dimensional wave-vector, must be evaluated as $\delta>0$ in
$z=\omega+i\delta$ tends to $0$. Depending on the values of $\mathbf{x}$ and
$\omega$ different results are obtained, it can be positive, negative or
imaginary. We find that reflective contributions to $\mathsf{G}(\mathbf{x,y}%
,\pm\hat{\omega}+i0)$ vanish in the radiative regime, $\hat{\omega}>\kappa$,
and transmission is also modified substantially. This confirms the results by
Pendry \cite{Pendry}. In the evanescent regime $\mathsf{G}(\mathbf{x,y},z)$
has poles in $z=\pm\hat{\omega}$, giving finite contributions, proportional to
$\exp[\pm i\hat{\omega}t]$, to the electric field $%
{\bm{E}}%
(\mathbf{x},t)$. It turns out that $\mathsf{K}$, the generator of the time
evolution in the AFF, has $\pm\hat{\omega}$ as eigenvalues with infinite
degeneracy, the latter giving rise to the above poles. Although we do not
discuss quantization, we note that this feature gives rise to an interesting
structure of the associated field Hamiltonian. In addition to the eigenvalue
$0$, associated with the vacuum state, now $\pm\hat{\omega}$ are also
eigenvalues. This affects radiative decay constants of excited atoms, as is
discussed in Section VII.

A word about notation: With a dispersive system we mean a dispersive,
non-absorptive system. Inner products are denoted as $(f,g)=\langle
g|f\rangle$. The unit vector along $\mathbf{a}\in\mathbb{R}^{3}$ is
$\mathbf{e}_{\mathbf{a}}=\mathbf{a}/a$, $a=|\mathbf{a}|$. The three Cartesian
axes are denoted by $X_{1}$, $X_{2}$ and $X_{3}$ with corresponding unit
vectors $\mathbf{e}_{1}$, $\mathbf{e}_{2}$ and $\mathbf{e}_{3}$. The component
of $\mathbf{a}\perp\mathbf{e}_{3}$ is denoted by $\mathbf{a}^{\perp}$.
\textsf{U} is the unit $3\times3$ matrix. Transposes of matrices are indicated
by means of the superscript $T$ and their Hermitean adjoints by \dag. Square
roots are defined in the usual way with non-negative imaginary part.
$\mathsf{I}_{\mathcal{A}}(x)$is the characteristic function for the set
$\mathcal{A}$, $\mathsf{I}_{\mathcal{A}}(x)=1$ for $x\in\mathcal{A}$ and
$\mathsf{I}_{\mathcal{A}}(x)=0$ for $x\notin\mathcal{A}$.

\section{Background}

Starting point is the set of linear phenomenological Maxwell's equations for
the case that permanent polarization and magnetization are absent (we set
$\varepsilon_{0}=\mu_{0}=1$ for brevity)%
\begin{align}
\partial_{t}%
{\bm{D}}%
(\mathbf{x},t)  &  =\partial_{\mathbf{x}}\times%
{\bm{H}}%
(\mathbf{x},t),\;\partial_{t}%
{\bm{B}}%
(\mathbf{x},t)=-\partial_{\mathbf{x}}\times%
{\bm{E}}%
(\mathbf{x},t),\nonumber\\
\partial_{\mathbf{x}}\cdot%
{\bm{D}}%
(\mathbf{x},t)  &  =0,\;\partial_{\mathbf{x}}\cdot%
{\bm{B}}%
(\mathbf{x},t)=0, \label{2.1}%
\end{align}
with the constitutive equations%
\begin{align}%
{\bm{D}}%
(\mathbf{x},t)  &  =%
{\bm{E}}%
(\mathbf{x},t)+%
{\bm{P}}%
(\mathbf{x},t),\;%
{\bm{P}}%
(\mathbf{x},t)=\int_{t_{0}}^{t}ds%
{\bm{\chi}}%
_{e}(\mathbf{x},t-s)\cdot%
{\bm{E}}%
(\mathbf{x},s),\nonumber\\%
{\bm{H}}%
(\mathbf{x},t)  &  =%
{\bm{B}}%
(\mathbf{x},t)-%
{\bm{M}}%
(\mathbf{x},t),\;%
{\bm{M}}%
(\mathbf{x},t)=\int_{t_{0}}^{t}ds%
{\bm{\chi}}%
_{m}(\mathbf{x},t-s)\cdot%
{\bm{H}}%
(\mathbf{x},s). \label{2.2}%
\end{align}
Here $%
{\bm{\chi}}%
_{e}(\mathbf{x},t)$ and $%
{\bm{\chi}}%
_{m}(\mathbf{x},t)$ are the electric and magnetic susceptibility tensors. We
also introduce the current densities $%
{\bm{J}}%
_{e}(\mathbf{x},t)=\partial_{t}%
{\bm{P}}%
(\mathbf{x},t)$ and $%
{\bm{J}}%
_{m}(\mathbf{x},t)=\partial_{t}%
{\bm{M}}%
(\mathbf{x},t)$. Causality requires that the susceptibilities $%
{\bm{\chi}}%
_{e}(\mathbf{x},t)$ and $%
{\bm{\chi}}%
_{m}(\mathbf{x},t)$ vanish for $t<0$. Assuming no initial surges in $%
{\bm{P}}%
(\mathbf{x},t)$ and $%
{\bm{M}}%
(\mathbf{x},t)$ at $t=t_{0}$, so $%
{\bm{J}}%
_{e}(\mathbf{x},t_{0})=%
{\bm{J}}%
_{m}(\mathbf{x},t_{0})=0$, we also have $%
{\bm{\chi}}%
_{e}(\mathbf{x},t_{0})=%
{\bm{\chi}}%
_{m}(\mathbf{x},t_{0})=0$. Indeed, currents are due to the motion of charged,
massive, particles and their velocity cannot be changed instantaneously. This
property is found in linear response expressions and also, for instance, for
the Lorentz case. Denoting $\partial_{t}%
{\bm{\chi}}%
(t)=%
{\bm{\chi}}%
^{\prime}(t)$, \ we then obtain
\begin{equation}%
{\bm{J}}%
_{e}(\mathbf{x},t)=\int_{t_{0}}^{t}ds%
{\bm{\chi}}%
_{e}^{\prime}(\mathbf{x},t-s)\cdot%
{\bm{E}}%
(\mathbf{x},s),\;%
{\bm{J}}%
_{m}(\mathbf{x},t)=\int_{t_{0}}^{t}ds%
{\bm{\chi}}%
_{m}^{\prime}(\mathbf{x},t-s)\cdot%
{\bm{H}}%
(\mathbf{x},s). \label{2.3}%
\end{equation}
As a matrix, $%
{\bm{\chi}}%
_{e,m}(\mathbf{x},t)$ are assumed to be symmetric (this property explicitly
holds for linear response expressions if the unperturbed matter Hamiltonian is
time-reversal invariant). \newline\textit{Remarks: }\newline1) The initial
time $t_{0}$ can have any value, in particular $t_{0}=-\infty$. However, in
view of the Laplace-transformed equations, introduced later on, $t_{0}=0$ is a
convenient choice. This case is often realized in practical situations. For
instance, in describing scattering of an electromagnetic wave-packet with
bounded support from a material object, the wave-packet is initially, as
$t\rightarrow-\infty$, well separated from the object and, in view of the
hyperbolic nature of Maxwell's equations, the support remains bounded and,
contrary to the Schr\"{o}dinger case, it takes a non-zero time for the support
to reach the object. Thus the polarization and magnetization vanish for times
smaller than some finite $t_{0}$, which we set equal to $0$.\newline2) It is
customary \cite{Jackson} to relate $%
{\bm{M}}%
$ to $%
{\bm{H}}%
$, rather than $%
{\bm{B}}%
$, although $%
{\bm{B}}%
$ is the more fundamental field. Indeed, interactions with atoms are in terms
of the microscopic vector potential $%
{\bm{A}}%
$, which is related to the microscopic $%
{\bm{B}}%
$-field. But note that if the particles are in vacuum, sufficiently far away
from the medium, the microscopic $%
{\bm{B}}%
$-field \ equals the macroscopic one and both equal $%
{\bm{H}}%
$ at the particle coordinates. We note further that linear response
expressions usually relate the magnetization to the (microscopic) $%
{\bm{B}}%
$-field.

Next we introduce the Fourier decomposition%
\begin{align}%
{\bm{\chi}}%
_{e,m}^{\prime}(\mathbf{x},t)  &  =\int d\lambda\exp[-i\lambda t]%
\bm{\nu}%
_{e,m}(\mathbf{x},\lambda),\nonumber\\%
\bm{\nu}%
_{e,m}(\mathbf{x},\lambda)  &  =\frac{1}{2\pi}\int dt\exp[i\lambda t]%
{\bm{\chi}}%
_{e,m}^{\prime}(\mathbf{x},t)\nonumber\\
&  =\frac{1}{2\pi}\int_{0}^{\infty}dt\exp[i\lambda t]%
{\bm{\chi}}%
_{e,m}^{\prime}(\mathbf{x},t). \label{2.4}%
\end{align}
Since $%
{\bm{\chi}}%
_{e,m}^{\prime}(\mathbf{x},t)$ are real, we have $%
\bm{\nu}%
_{e,m}(\mathbf{x},-\lambda)=\overline{%
\bm{\nu}%
_{e,m}(\mathbf{x},\lambda)}$ in the sense that this relation holds for each
component of these tensors. We also assume that the system is passive. This
means that the electromagnetic energy%
\begin{equation}
\mathcal{E}_{em}(t)=\dfrac{1}{2}\int d\mathbf{x}\{%
{\bm{E}}%
(\mathbf{x},t)^{2}+%
{\bm{H}}%
(\mathbf{x},t)^{2}\}, \label{2.5}%
\end{equation}
cannot increase as a function of time. So initial population inversions in the
material system are excluded. Then%
\begin{align}
\mathcal{E}_{em}(t)-\mathcal{E}_{em}(t_{0})  &  =\int_{t_{0}}^{t}%
ds\partial_{s}\mathcal{E}_{em}(s)=-\int_{t_{0}}^{t}ds\int d\mathbf{x\{}%
{\bm{J}}%
_{e}(\mathbf{x},s)\cdot%
{\bm{E}}%
(\mathbf{x},s)+%
{\bm{J}}%
_{m}(\mathbf{x},s)\cdot%
{\bm{H}}%
(\mathbf{x},s)\}\nonumber\\
&  =-\int_{t_{0}}^{t}ds\int_{t_{0}}^{s}du\int d\mathbf{x\{}%
{\bm{\chi}}%
_{e}^{\prime}(\mathbf{x},s-u)%
{\bm{:}}%
{\bm{E}}%
(\mathbf{x},u)%
{\bm{E}}%
(\mathbf{x},s)+%
{\bm{\chi}}%
_{m}^{\prime}(\mathbf{x},s-u)%
{\bm{:}}%
{\bm{H}}%
(\mathbf{x},u)%
{\bm{H}}%
(\mathbf{x},s)\}\nonumber\\
&  =-\int_{t_{0}}^{t}ds\int_{t_{0}}^{t}du\int d\mathbf{x\{}%
{\bm{\chi}}%
_{e}^{\prime}(\mathbf{x},s-u)%
{\bm{:}}%
{\bm{E}}%
(\mathbf{x},u)%
{\bm{E}}%
(\mathbf{x},s)+%
{\bm{\chi}}%
_{m}^{\prime}(\mathbf{x},s-u)%
{\bm{:}}%
{\bm{H}}%
(\mathbf{x},u)%
{\bm{H}}%
(\mathbf{x},s)\}\nonumber\\
&  =-\int d\mathbf{x[}%
\bm{\nu}%
_{e}(\mathbf{x},\lambda)%
{\bm{:}}%
\{\int_{t_{0}}^{t}du\exp[i\lambda u]%
{\bm{E}}%
(\mathbf{x},u)\}\{\int_{t_{0}}^{t}ds\exp[-i\lambda s]%
{\bm{E}}%
(\mathbf{x},s)\}\nonumber\\
&  +%
\bm{\nu}%
_{m}(\mathbf{x},\lambda)%
{\bm{:}}%
\{\int_{t_{0}}^{t}du\exp[i\lambda u]%
{\bm{H}}%
(\mathbf{x},u)\}\{\int_{t_{0}}^{t}ds\exp[-i\lambda s]%
{\bm{H}}%
(\mathbf{x},s)\}]\leqq0, \label{2.6}%
\end{align}
so%
\begin{equation}%
\bm{\nu}%
_{e,m}(\mathbf{x},\lambda)\geqslant0. \label{2.7}%
\end{equation}
This result also emerges in linear response expressions if the initial density
operator for the material system is a function of its Hamiltonian and the
level population decreases with increasing energy as is the case for a
canonical distribution. Now $%
\bm{\nu}%
_{e,m}(\mathbf{x},-\lambda)=%
\bm{\nu}%
_{e,m}(\mathbf{x},\lambda)\geqslant0$, leading to%
\begin{align}%
{\bm{\chi}}%
_{e}(\mathbf{x},t)  &  =\int d\lambda\dfrac{\sin(\lambda t)}{\lambda}%
\bm{\nu}%
_{e}(\mathbf{x},\lambda),\;%
{\bm{\chi}}%
_{m}(\mathbf{x},t)=\int d\lambda\dfrac{\sin(\lambda t)}{\lambda}%
\bm{\nu}%
_{m}(\mathbf{x},\lambda),\nonumber\\%
{\bm{\chi}}%
_{e}^{\prime}(\mathbf{x},t)  &  =\int d\lambda\cos(\lambda t)%
\bm{\nu}%
_{e}(\mathbf{x},\lambda),\;%
{\bm{\chi}}%
_{m}^{\prime}(\mathbf{x},t)=\int d\lambda\cos(\lambda t)%
\bm{\nu}%
_{m}(\mathbf{x},\lambda). \label{2.8}%
\end{align}
We introduce Laplace transforms according to ($\operatorname{Im}z>0$)%
\begin{equation}
\hat{f}(z)=\int_{0}^{\infty}dt\exp[izt]f(t),\;f(t)=\dfrac{1}{2\pi}\int
_{\Gamma}dz\exp[-izt]\hat{f}(z), \label{2.9}%
\end{equation}
where $\Gamma$ is a path running from $-\infty$ to $+\infty$ at some distance
$\delta>0$ parallel to the real axis. Then, for $t_{0}\leqslant0$, performing
a partial integration and using $%
{\bm{\chi}}%
_{e,m}(t)=0$, $t<0$,
\begin{align}
&
{\hat{\bm{\chi}}}%
_{e,m}(\mathbf{x},z)=\int_{t_{0}}^{\infty}dt\exp[izt]%
{\bm{\chi}}%
_{e,m}(\mathbf{x},t)=\int_{0}^{\infty}dt\exp[izt]%
{\bm{\chi}}%
_{e,m}(\mathbf{x},t)\nonumber\\
&  =-\frac{1}{iz}\int_{0}^{\infty}dt\exp[izt]%
{\bm{\chi}}%
_{e,m}^{\prime}(\mathbf{x},t)=-\frac{1}{iz}\int_{0}^{\infty}dt\exp[izt]\int
d\lambda%
\bm{\nu}%
_{e,m}(\mathbf{x},\lambda)\exp[-i\lambda t]=\int d\lambda%
\bm{\nu}%
_{e,m}(\mathbf{x},\lambda)\frac{1}{\lambda^{2}-z^{2}}. \label{2.10}%
\end{align}
Causality and passivity imply that these are the most general expressions for
$%
{\hat{\bm{\chi}}}%
_{e,m}(\mathbf{x},z)$ \cite{Drude-Lorentz} (the latter, being analytic in the
upper half plane, are so-called Herglotz functions, which can always be
represented in the above form). In general $%
\bm{\nu}%
_{e,m}(\mathbf{x},\lambda)$ is made up of \ integrable functions, leading to
absorptive systems, and of $\delta$-function contributions. The second give
rise to a set of dispersive Lorentz terms. In fact we can say that causal,
passive, dispersive susceptibilities consist of a (possibly infinite) sum of
dispersive Lorentz terms \cite{Drude-Lorentz}.

\noindent Then, in the isotropic, dispersive, case, $%
{\hat{\bm{\chi}}}%
(\mathbf{x},z)=\hat{\chi}(\mathbf{x},z)\mathsf{U},$%

\begin{equation}
\nu_{e,m}(\mathbf{x},\lambda)=\sum_{n}\Omega_{e,m}^{n}(\mathbf{x})^{2}%
\delta(\lambda-\lambda_{e,m}^{n}),\;\hat{\chi}_{e,m}(\mathbf{x},z)=-\sum
_{n}\frac{\Omega_{e,m}^{n}(\mathbf{x})^{2}}{z^{2}-(\lambda_{e,m}^{n})^{2}}.
\label{2.11}%
\end{equation}
Here we recall that, disregarding $\mathbf{x}$-dependencies, an absorptive
Lorentz contribution is given by%
\begin{equation}
\hat{\chi}_{\gamma}(z)=-\frac{\Omega^{2}}{z^{2}+i\gamma z-\omega_{0}^{2}%
},\;\chi_{\gamma}(t)=\Omega^{2}\exp[-\frac{\gamma t}{2}]\frac{\sin\omega_{1}%
t}{\omega_{1}},\;\omega_{1}=\sqrt{\omega_{0}^{2}+\frac{\gamma^{2}}{4}%
},\;\Omega,\omega_{0},\gamma>0. \label{2.12}%
\end{equation}
For $\omega\in\mathbb{R}$,%
\begin{align}
\hat{\chi}_{e,m}(\mathbf{x},\omega)  &  =\int_{0}^{\infty}dt\exp[i\omega
t]\chi_{e,m}(t)=\frac{1}{\omega}\int d\lambda\frac{1}{\lambda-\omega-i0}%
\nu_{e,m}(\mathbf{x},\lambda),\nonumber\\
\operatorname{Im}\hat{\chi}_{e,m}(\mathbf{x},\omega)  &  =\frac{\pi}{\omega
}\nu_{e,m}(\mathbf{x},\omega). \label{2.13}%
\end{align}
In the dispersive case, $\gamma=0$,
\begin{equation}
\hat{\chi}(\omega+i0)=-\frac{\Omega^{2}}{\omega^{2}-\omega_{0}^{2}}+i\pi
\delta(\omega^{2}-\omega_{0}^{2}),\;\chi(t)=\Omega^{2}\frac{\sin\omega_{0}%
t}{\omega_{0}}. \label{2.14}%
\end{equation}
Here the $\delta$-function insures the validity of the Kramers-Kronig
relations although it does not play a further role \cite{Drude-Lorentz}.
\newline Negative index systems are characterized by negative real
permeabilities for some real frequencies. In the Lorentz case this can happen
at a few discrete values of $z$. Indeed, for a single dispersive Lorentz term,
setting%
\begin{equation}
1+\hat{\chi}(z)=-1, \label{2.15}%
\end{equation}
we find%
\begin{equation}
z=\pm\hat{\omega},\;\hat{\omega}=\sqrt{\omega_{0}^{2}+\frac{\Omega^{2}}{2}},
\label{2.16}%
\end{equation}
but if absorption is present, only complex solutions with negative imaginary
part are obtained.

\section{Time evolution}

In this section we extend the AFF to include magnetization. Let $\mathcal{A}%
\subset\mathbb{R}^{3}$ be the set containing the medium, so $%
\bm{\nu}%
_{e,m}(\mathbf{x},\lambda)$ vanish if $\mathbf{x}\notin\mathcal{A}$ as do $%
{\bm{F}}%
_{2}$, $%
{\bm{F}}%
_{3}$, $%
{\bm{F}}%
_{5}$ and $%
{\bm{F}}%
_{6}$ below. In \cite{LAD} $%
{\bm{E}}%
$ and $%
{\bm{B}}%
$ were used as the electromagnetic components. It is possible to do so in the
present situation, but choosing $%
{\bm{E}}%
$\ and $%
{\bm{H}}%
$ gives somewhat more symmetric formulae. We introduce%
\begin{align}%
{\bm{F}}%
_{1}(\mathbf{x},t)  &  =%
{\bm{E}}%
(\mathbf{x},t),\nonumber\\%
{\bm{F}}%
_{2}(\mathbf{x},\lambda,t)  &  =\mathsf{I}_{\mathcal{A}}(\mathbf{x}%
)\int_{t_{0}}^{t}ds\sin\{\lambda(t-s)\}%
{\bm{E}}%
(\mathbf{x},s),\nonumber\\%
{\bm{F}}%
_{3}(\mathbf{x},\lambda,t)  &  =\mathsf{I}_{\mathcal{A}}(\mathbf{x}%
)\int_{t_{0}}^{t}ds\cos\{\lambda(t-s)\}%
{\bm{H}}%
(\mathbf{x},s),\nonumber\\%
{\bm{F}}%
_{4}(\mathbf{x},t)  &  =%
{\bm{H}}%
(\mathbf{x},t),\nonumber\\%
{\bm{F}}%
_{\ 5}(\mathbf{x},\lambda,t)  &  =\mathsf{I}_{\mathcal{A}}(\mathbf{x}%
)\int_{t_{0}}^{t}ds\sin\{\lambda(t-s)\}%
{\bm{H}}%
(\mathbf{x},s),\nonumber\\%
{\bm{F}}%
_{6}(\mathbf{x},\lambda,t)  &  =\mathsf{I}_{\mathcal{A}}(\mathbf{x}%
)\int_{t_{0}}^{t}ds\cos\{\lambda(t-s)\}%
{\bm{E}}%
(\mathbf{x},s). \label{3.1}%
\end{align}
Then%
\begin{align}
&
{\bm{J}}%
_{e}(\mathbf{x},t)=\int d\lambda%
\bm{\nu}%
_{e}(\mathbf{x},\lambda)\cdot\int_{t_{0}}^{t}ds\cos\{\lambda(t-s)\}%
{\bm{F}}%
_{1}(\mathbf{x},s)=\int d\lambda%
\bm{\nu}%
_{e}(\mathbf{x},\lambda)\cdot%
{\bm{F}}%
_{\ 6}(\mathbf{x},\lambda,s),\nonumber\\
&
{\bm{J}}%
_{m}(\mathbf{x},t)=\int d\lambda%
\bm{\nu}%
_{m}(\mathbf{x},\lambda)\cdot\int_{t_{0}}^{t}ds\cos\{\lambda(t-s)\}%
{\bm{F}}%
_{4}(\mathbf{x},s)=\int d\lambda%
\bm{\nu}%
_{m}(\mathbf{x},\lambda)\cdot%
{\bm{F}}%
_{3}(\mathbf{x},\lambda,s), \label{3.2}%
\end{align}
and%
\begin{align}
\partial_{t}%
{\bm{F}}%
_{1}(\mathbf{x},t)  &  =\partial_{\mathbf{x}}\times%
{\bm{F}}%
_{4}(\mathbf{x},t)-\int d\lambda%
\bm{\nu}%
_{e}(\mathbf{x},\lambda)\cdot%
{\bm{F}}%
_{6}(\mathbf{x},\lambda,t),\nonumber\\
\partial_{t}%
{\bm{F}}%
_{2}(\mathbf{x},\lambda,t)  &  =\lambda%
{\bm{F}}%
_{6}(\mathbf{x},\lambda,t),\nonumber\\
\partial_{t}%
{\bm{F}}%
_{3}(\mathbf{x},\lambda,t)  &  =\mathsf{I}_{\mathcal{A}}(\mathbf{x})%
{\bm{F}}%
_{4}(\mathbf{x},t)-\lambda%
{\bm{F}}%
_{5}(\mathbf{x},\lambda,t),\nonumber\\
\partial_{t}%
{\bm{F}}%
_{4}(\mathbf{x},t)  &  =-\partial_{\mathbf{x}}\times%
{\bm{F}}%
_{1}(\mathbf{x},t)-\int d\lambda%
\bm{\nu}%
_{m}(\mathbf{x},\lambda)\cdot%
{\bm{F}}%
_{3}(\mathbf{x},\lambda,t),\nonumber\\
\partial_{t}%
{\bm{F}}%
_{5}(\mathbf{x},\lambda,t)  &  =\lambda%
{\bm{F}}%
_{3}(\mathbf{x},\lambda,t),\nonumber\\
\partial_{t}%
{\bm{F}}%
_{6}(\mathbf{x},\lambda,t)  &  =\mathsf{I}_{\mathcal{A}}(\mathbf{x})%
{\bm{F}}%
_{1}(\mathbf{x},t)-\lambda%
{\bm{F}}%
_{2}(\mathbf{x},\lambda,t). \label{3.3}%
\end{align}
Note that%
\begin{equation}%
{\bm{F}}%
_{2,3,5,6}(\mathbf{x},\lambda,t_{0})=0. \label{3.4}%
\end{equation}
In condensed notation,%
\begin{equation}
\partial_{t}%
{\bm{F}}%
(t)=-i\mathsf{K}\cdot%
{\bm{F}}%
(t). \label{3.5}%
\end{equation}
Let $\langle\lambda|%
\bm{\nu}%
_{e,m}\rangle=%
\bm{\nu}%
_{e,m}(\mathbf{x},\lambda)$, $%
\bm{\epsilon}%
$ be the Levi-Civita symbol and $\mathbf{p}=-i\partial_{\mathbf{x}}$ so $($ $%
\bm{\epsilon}%
\cdot\mathbf{p})\cdot\mathbf{f}=i\partial_{\mathbf{x}}\times\mathbf{f}$). Then%
\begin{equation}%
{\bm{F}}%
=\left(
\begin{array}
[c]{c}%
{\bm{F}}%
_{1}\\%
{\bm{F}}%
_{2}\\%
{\bm{F}}%
_{3}\\%
{\bm{F}}%
_{4}\\%
{\bm{F}}%
_{5}\\%
{\bm{F}}%
_{6}%
\end{array}
\right)  ,\;\mathsf{K}=\left(
\begin{array}
[c]{cccccc}%
0 & 0 & 0 &
\bm{\epsilon}%
\cdot\mathbf{p} & 0 & -i\langle%
\bm{\nu}%
_{e}|\\
0 & 0 & 0 & 0 & 0 & i\lambda\\
0 & 0 & 0 & i\mathsf{I}_{\mathcal{A}}(\mathbf{x}) & -i\lambda & 0\\
-%
\bm{\epsilon}%
\cdot\mathbf{p} & 0 & -i\langle%
\bm{\nu}%
_{m}| & 0 & 0 & 0\\
0 & 0 & i\lambda & 0 & 0 & 0\\
i\mathsf{I}_{\mathcal{A}}(\mathbf{x}) & -i\lambda & 0 & 0 & 0 & 0
\end{array}
\right)  =\left(
\begin{array}
[c]{cc}%
0 & \mathsf{K}_{em}\\
\mathsf{K}_{me} & 0
\end{array}
\right)  . \label{3.6}%
\end{equation}
Thus, as in the dielectric case, \textsf{K} is symplectic. Let now%
\begin{align}
&  \mathcal{E}_{e}(t)=\dfrac{1}{2}\int d\mathbf{x}\int d\lambda%
\bm{\nu}%
_{e}(\mathbf{x},\lambda)%
{\bm{:}}%
\{%
{\bm{F}}%
_{2}(\mathbf{x},\lambda,t)%
{\bm{F}}%
_{2}(\mathbf{x},\lambda,t)+%
{\bm{F}}%
_{6}(\mathbf{x},\lambda,t)%
{\bm{F}}%
_{6}(\mathbf{x},\lambda,t)\},\nonumber\\
&  \mathcal{E}_{m}(t)=\dfrac{1}{2}\int d\mathbf{x}\int d\lambda%
\bm{\nu}%
_{m}(\mathbf{x},\lambda)%
{\bm{:}}%
\{%
{\bm{F}}%
_{3}(\mathbf{x},\lambda,t)%
{\bm{F}}%
_{3}(\mathbf{x},\lambda,t)+%
{\bm{F}}%
_{5}(\mathbf{x},\lambda,t)%
{\bm{F}}%
_{5}(\mathbf{x},\lambda,t)\}. \label{3.7}%
\end{align}
Then, with $\mathcal{E}_{em}(t)$ as given by Eq. (2.5),
\begin{align}
&  \partial_{t}\mathcal{E}_{em}(t)=-\int d\mathbf{x\{}%
{\bm{J}}%
_{e}(\mathbf{x},t)\cdot%
{\bm{F}}%
_{1}(\mathbf{x},t)+%
{\bm{J}}%
_{m}(\mathbf{x},t)\cdot%
{\bm{F}}%
_{4}(\mathbf{x},t)\},\nonumber\\
&  \partial_{t}\mathcal{E}_{e}(t)=\int d\mathbf{x}\int d\lambda%
\bm{\nu}%
_{e}(\mathbf{x},\lambda)%
{\bm{:}}%
{\bm{F}}%
_{6}(\mathbf{x},\lambda,t)%
{\bm{F}}%
_{1}(\mathbf{x},t)=\int d\mathbf{x}%
{\bm{J}}%
_{e}(\mathbf{x},t)\cdot%
{\bm{F}}%
_{1}(\mathbf{x},t),\nonumber\\
&  \partial_{t}\mathcal{E}_{m}(t)=\int d\mathbf{x}\int d\lambda%
\bm{\nu}%
_{m}(\mathbf{x},\lambda)%
{\bm{:}}%
{\bm{F}}%
_{5}(\mathbf{x},\lambda,t)%
{\bm{F}}%
_{4}(\mathbf{x},t)=\int d\mathbf{x}%
{\bm{J}}%
_{m}(\mathbf{x},t)\cdot%
{\bm{F}}%
_{4}(\mathbf{x},t), \label{3.8}%
\end{align}
so, with%
\begin{equation}
\mathcal{E}(t)=\mathcal{E}_{em}(t)+\mathcal{E}_{e}(t)+\mathcal{E}_{m}(t),
\label{3.9}%
\end{equation}
we have%
\begin{equation}
\partial_{t}\mathcal{E}(t)=0, \label{3.10}%
\end{equation}
i.e., $\mathcal{E}(t)$ is conserved in time. At this point we note that the
standard expression for the conserved energy is%
\begin{equation}
\partial_{t}\mathcal{E}=\int d\mathbf{x}\{%
{\bm{E}}%
(\mathbf{x},t)\cdot\partial_{t}%
{\bm{D}}%
(\mathbf{x},t)+%
{\bm{H}}%
(\mathbf{x},t)\cdot\partial_{t}%
{\bm{B}}%
(\mathbf{x},t)\}=\partial_{t}\mathcal{E}_{em}(t)+\int d\mathbf{x}\{%
{\bm{E}}%
(\mathbf{x},t)\cdot%
{\bm{J}}%
_{e}(\mathbf{x},t)+%
{\bm{H}}%
(\mathbf{x},t)\cdot%
{\bm{J}}%
_{m}(\mathbf{x},t)\}, \label{3.11}%
\end{equation}
so the two expressions agree.\medskip\newline We introduce the inner product%
\begin{align}
(%
{\bm{F}}%
,%
{\bm{G}}%
)  &  =\int d\mathbf{x[}%
{\bm{F}}%
_{1}(\mathbf{x})\cdot\overline{%
{\bm{G}}%
_{1}(\mathbf{x})}+%
{\bm{F}}%
_{4}(\mathbf{x})\cdot\overline{%
{\bm{G}}%
_{4}(\mathbf{x})}]+\int_{\mathcal{A}}d\mathbf{x}\int d\lambda%
\bm{\nu}%
_{e}(\mathbf{x},\lambda)%
{\bm{:}}%
[%
{\bm{F}}%
_{2}(\mathbf{x},\lambda)\overline{%
{\bm{G}}%
_{2}(\mathbf{x},\lambda)}\nonumber\\
&  +%
{\bm{F}}%
_{6}(\mathbf{x},\lambda)\overline{%
{\bm{G}}%
_{6}(\mathbf{x},\lambda)}]+\int_{\mathcal{A}}d\mathbf{x}\int d\lambda%
\bm{\nu}%
_{m}(\mathbf{x},\lambda)%
{\bm{:}}%
[%
{\bm{F}}%
_{3}(\mathbf{x},\lambda)\overline{%
{\bm{G}}%
_{3}(\mathbf{x},\lambda)}+%
{\bm{F}}%
_{5}(\mathbf{x},\lambda)\overline{%
{\bm{G}}%
_{5}(\mathbf{x},\lambda)}], \label{3.12}%
\end{align}
which defines the Hilbert space%
\begin{align}
\mathcal{K}  &  =\oplus_{j=1}^{6}\mathcal{K}_{j},\;\mathcal{K}_{1}%
=\mathcal{K}_{4}=L^{2}(\mathbb{R}^{3},d\mathbf{x};\mathbb{C}^{3}),\nonumber\\
\mathcal{K}_{2}  &  =\mathcal{K}_{6}=L^{2}(\mathcal{A},d\mathbf{x}%
;\mathbb{C}^{3})\otimes L^{2}(\mathbb{R},%
\bm{\nu}%
_{e}d\lambda),\;\mathcal{K}_{3}=\mathcal{K}_{5}=L^{2}(\mathcal{A}%
,d\mathbf{x};\mathbb{C}^{3})\otimes L^{2}(\mathbb{R},%
\bm{\nu}%
_{m}d\lambda). \label{3.13}%
\end{align}
Then
\begin{equation}
\parallel%
{\bm{F}}%
(t)\parallel^{2}=(%
{\bm{F}}%
(t),%
{\bm{F}}%
(t))=2\mathcal{E}, \label{3.14}%
\end{equation}
is conserved in time. In the Appendix it is shown (for simpler notation the
isotropic case is considered) that $\mathsf{K}$ is selfadjoint in\textsf{
}$\mathcal{K}$ under some mild conditions on the susceptibilities ( $|%
{\bm{\chi}}%
_{e,m}^{\prime}(\mathbf{x},0)|\leqslant c<\infty$), which we assume to hold
from now on. In fact in
\begin{equation}
\mathsf{K}=\mathsf{K}_{0}+\mathsf{K}_{1}, \label{3.15}%
\end{equation}
where%
\begin{equation}
\mathsf{K}_{0}=\left(
\begin{array}
[c]{cccccc}%
0 & 0 & 0 &
\bm{\epsilon}%
\cdot\mathbf{p} & 0 & 0\\
0 & 0 & 0 & 0 & 0 & i\lambda\\
0 & 0 & 0 & 0 & -i\lambda & 0\\
-%
\bm{\epsilon}%
\cdot\mathbf{p} & 0 & 0 & 0 & 0 & 0\\
0 & 0 & i\lambda & 0 & 0 & 0\\
0 & -i\lambda & 0 & 0 & 0 & 0
\end{array}
\right)  ,\;\mathsf{K}_{1}=\left(
\begin{array}
[c]{cccccc}%
0 & 0 & 0 & 0 & 0 & -i\langle%
\bm{\nu}%
_{e}|\\
0 & 0 & 0 & 0 & 0 & 0\\
0 & 0 & 0 & i & 0 & 0\\
0 & 0 & -i\langle%
\bm{\nu}%
_{m}| & 0 & 0 & 0\\
0 & 0 & 0 & 0 & 0 & 0\\
i & 0 & 0 & 0 & 0 & 0
\end{array}
\right)  . \label{3.16}%
\end{equation}
$\mathsf{K}_{1}$ is a bounded selfadjoint operator. Thus we are dealing with a
unitary time evolution on $\mathcal{K},$%
\begin{equation}
\mathsf{W}(t)=\exp[-i\mathsf{K}t]. \label{3.17}%
\end{equation}
As mentioned before it implies that the time evolution of the electromagnetic
fields is properly defined. In case the electromagnetic fields are square
integrable at the initial time, this remains true at all later times (note the
passivity condition above). This is not evident in the original formulation
which contains time convolutions.\medskip\newline\textit{Remark:} Note that
passivity is not required to obtain the conservation of $\mathcal{E}(t)$.
Since $\mathcal{E}_{em}(0)\geqslant0$ and $\mathcal{E}_{e}(0)=\mathcal{E}%
_{m}(0)=0$ we still have $\mathcal{E}(t)=\mathcal{E}(0)\geqslant0$. However,
without it, $\mathcal{E}_{em}(t)$ may increase in time and $\mathcal{E}%
_{e}(t)$ and $\mathcal{E}_{m}(t)$ become negative for $t>0$ ($%
\bm{\nu}%
_{e,m}(\mathbf{x},\lambda)$ may no longer be non-negative). We still can
introduce the inner product $(%
{\bm{F}}%
,%
{\bm{G}}%
)$ but the associated norm is also no longer non-negative definite.\medskip
\newline For dispersive systems $%
\bm{\nu}%
_{e,m}(\mathbf{x},\lambda)$ becomes a sum of $\delta$-functions. Let us assume
that only one dispersive Lorentz contribution is present in both $\chi$'s and
that the medium is homogeneous and isotropic over $\mathcal{A}$. Then
$\nu_{e,m}(\mathbf{x})=\nu_{e,m}\mathsf{I}_{\mathcal{A}}(\mathbf{x})$, so in
$\mathcal{K}_{2,3,5,6}$ the $\mathbf{x}$- integration is over $\mathcal{A}$
and
\begin{align}
\chi_{e}(\mathbf{x},t)  &  =\Omega_{e}^{2}\mathsf{I}_{\mathcal{A}}%
(\mathbf{x})\frac{\sin\lambda_{e}t}{\lambda_{e}},\;\chi_{e}^{\prime
}(\mathbf{x},t)=\Omega_{e}^{2}\mathsf{I}_{\mathcal{A}}(\mathbf{x})\cos
\lambda_{e}t,\nonumber\\
\chi_{m}(\mathbf{x},t)  &  =\Omega_{m}^{2}\mathsf{I}_{\mathcal{A}}%
(\mathbf{x})\frac{\sin\lambda_{m}t}{\lambda_{m}},\;\chi_{m}^{\prime
}(\mathbf{x},t)=\Omega_{m}^{2}\mathsf{I}_{\mathcal{A}}(\mathbf{x})\cos
\lambda_{m}t,\nonumber\\
\hat{\chi}_{e}(\mathbf{x},z)  &  =\frac{\Omega_{e}^{2}}{\lambda_{e}^{2}-z^{2}%
}\mathsf{I}_{\mathcal{A}}(\mathbf{x}),\;\hat{\chi}_{m}(\mathbf{x}%
,z)=\frac{\Omega_{m}^{2}}{\lambda_{m}^{2}-z^{2}}\mathsf{I}_{\mathcal{A}%
}(\mathbf{x}), \label{3.18}%
\end{align}
so
\begin{align}%
{\bm{J}}%
_{e}(\mathbf{x},t)  &  =\Omega_{e}^{2}\mathsf{I}_{\mathcal{A}}(\mathbf{x}%
)\int_{t_{0}}^{t}ds\cos\{\lambda_{e}(t-s)\}%
{\bm{E}}%
(\mathbf{x},s)\nonumber\\%
{\bm{J}}%
_{m}(\mathbf{x},t)  &  =\Omega_{m}^{2}\mathsf{I}_{\mathcal{A}}(\mathbf{x}%
)\int_{t_{0}}^{t}ds\cos\{\lambda_{m}(t-s)\}%
{\bm{H}}%
(\mathbf{x},s). \label{3.19}%
\end{align}
In this case, with $%
{\bm{F}}%
_{1}=%
{\bm{E}}%
$, $%
{\bm{F}}%
_{4}=%
{\bm{H}}%
$ and%
\begin{align}%
{\bm{F}}%
_{2}(\mathbf{x},t)  &  =\mathsf{I}_{\mathcal{A}}(\mathbf{x})\int_{t_{0}}%
^{t}ds\sin\{\lambda_{e}(t-s)\}%
{\bm{E}}%
(\mathbf{x},s),\;%
{\bm{F}}%
_{3}(\mathbf{x},t)=\mathsf{I}_{\mathcal{A}}(\mathbf{x})\int_{t_{0}}^{t}%
ds\cos\{\lambda_{m}(t-s)\}%
{\bm{H}}%
(\mathbf{x},s),\nonumber\\%
{\bm{F}}%
_{5}(\mathbf{x},t)  &  =\mathsf{I}_{\mathcal{A}}(\mathbf{x})\int_{t_{0}}%
^{t}ds\sin\{\lambda_{m}(t-s)\}%
{\bm{H}}%
(\mathbf{x},s),\;%
{\bm{F}}%
_{6}(\mathbf{x},t)=\mathsf{I}_{\mathcal{A}}(\mathbf{x})\int_{t_{0}}^{t}%
ds\cos\{\lambda_{e}(t-s)\}%
{\bm{E}}%
(\mathbf{x},s), \label{3.20}%
\end{align}
once more,%
\begin{equation}
\partial_{t}%
{\bm{F}}%
(t)=-i\mathsf{K}\cdot%
{\bm{F}}%
(t), \label{3.21}%
\end{equation}
where now%
\begin{equation}
\mathsf{K}=\left(
\begin{array}
[c]{cccccc}%
0 & 0 & 0 &
\bm{\epsilon}%
\cdot\mathbf{p} & 0 & -i\Omega_{e}^{2}\\
0 & 0 & 0 & 0 & 0 & i\lambda_{e}\\
0 & 0 & 0 & i\mathsf{I}_{\mathcal{A}}(\mathbf{x}) & -i\lambda_{m} & 0\\
-%
\bm{\epsilon}%
\cdot\mathbf{p} & 0 & -i\Omega_{m}^{2} & 0 & 0 & 0\\
0 & 0 & i\lambda_{m} & 0 & 0 & 0\\
i\mathsf{I}_{\mathcal{A}}(\mathbf{x}) & -i\lambda_{e} & 0 & 0 & 0 & 0
\end{array}
\right)  , \label{3.22}%
\end{equation}
whereas $\mathcal{K}_{2,6}$ and $\mathcal{K}_{3,5}$ reduce to $L^{2}%
(\mathcal{A},\Omega_{e}^{2}d\mathbf{x;}\mathbb{C}^{3})$ and $L^{2}%
(\mathcal{A},\Omega_{m}^{2}d\mathbf{x;}\mathbb{C}^{3})$, respectively. In case
there are more dispersive Lorentz terms in the susceptibilities the number of
auxiliary fields increases accordingly.

\section{Laplace-transformed fields}

The equations of motion can equivalently be expressed in terms of Laplace
transforms. Setting $t_{0}=0$, we obtain%
\begin{equation}%
{\hat{\bm{F}}}%
(z)=i[z-\mathsf{K}]^{-1}\cdot%
{\bm{F}}%
(0)=i\mathsf{R}(z)\cdot%
{\bm{F}}%
(0),\;\operatorname{Im}z>0. \label{4.1}%
\end{equation}
From this the relations for the various components of $%
{\hat{\bm{F}}}%
(z)$ can be obtained in terms of those of $%
{\bm{F}}%
(0)$ by projecting upon the appropriate subspace, see the Appendix. However, a
direct approach involves less calculations. Thus,%
\begin{align}
-iz%
{\hat{\bm{D}}}%
(\mathbf{x},z)-%
{\bm{D}}%
(\mathbf{x},0)  &  =-i(%
\bm{\epsilon}%
\cdot\mathbf{p)}\cdot%
{\hat{\bm{H}}}%
(\mathbf{x},z),\nonumber\\
-iz%
{\hat{\bm{B}}}%
(\mathbf{x},z)-%
{\bm{B}}%
(\mathbf{x},0)  &  =+i(%
\bm{\epsilon}%
\cdot\mathbf{p)}\cdot%
{\hat{\bm{E}}}%
(\mathbf{x},z). \label{4.2}%
\end{align}
Since%
\begin{equation}%
{\hat{\bm{D}}}%
(\mathbf{x},z)=%
{\bm{\varepsilon}}%
(\mathbf{x},z)\cdot%
{\hat{\bm{E}}}%
(\mathbf{x},z),\;%
{\hat{\bm{B}}}%
(\mathbf{x},z)=%
\bm{\mu}%
(\mathbf{x},z)\cdot%
{\hat{\bm{H}}}%
(\mathbf{x},z). \label{4.3}%
\end{equation}
we obtain, noting that $%
{\bm{D}}%
(\mathbf{x},0)=%
{\bm{E}}%
(\mathbf{x},0)$ and $%
{\bm{H}}%
(\mathbf{x},0)=%
{\bm{B}}%
(\mathbf{x},0)$,
\begin{equation}
\mathsf{L}^{e}(z)\cdot%
{\hat{\bm{E}}}%
(\mathbf{x},z)=\mathbf{g}^{e}(\mathbf{x},z),\;\mathsf{L}^{m}(z)\cdot%
{\hat{\bm{H}}}%
(\mathbf{x},z)=\mathbf{g}^{m}(\mathbf{x},z), \label{4.4}%
\end{equation}
where
\begin{align}
\mathsf{L}^{e}(z)  &  =z^{2}%
{\bm{\varepsilon}}%
(\mathbf{x},z)+(%
\bm{\epsilon}%
\cdot\mathbf{p)}\cdot%
\bm{\mu}%
(\mathbf{x},z)^{-1}\cdot(%
\bm{\epsilon}%
\cdot\mathbf{p}),\nonumber\\
\mathsf{L}^{m}(z)  &  =z^{2}%
\bm{\mu}%
(\mathbf{x},z)+(%
\bm{\epsilon}%
\cdot\mathbf{p)}\cdot%
{\bm{\varepsilon}}%
(\mathbf{x},z)^{-1}\cdot(%
\bm{\epsilon}%
\cdot\mathbf{p)},\nonumber\\
\mathbf{g}^{e}(\mathbf{x},z)  &  =iz%
{\bm{E}}%
(\mathbf{x},0)+i(%
\bm{\epsilon}%
\cdot\mathbf{p)}\cdot\{%
\bm{\mu}%
(\mathbf{x},z)^{-1}\cdot%
{\bm{H}}%
(\mathbf{x},0)\},\nonumber\\
\mathbf{g}^{m}(\mathbf{x},z)  &  =iz%
{\bm{H}}%
(\mathbf{x},0)-i(%
\bm{\epsilon}%
\cdot\mathbf{p)}\cdot\{%
{\bm{\varepsilon}}%
(\mathbf{x},z)^{-1}\cdot%
{\bm{E}}%
(\mathbf{x},0)\}. \label{4.5}%
\end{align}
Here $\mathsf{L}^{e}(z)$ and $\mathsf{L}^{m}(z)$ are the electric and magnetic
Helmholtz operators. Let now%
\begin{equation}
\mathsf{R}^{e}(z)=\mathsf{L}^{e}(z)^{-1},\;\mathsf{R}^{m}(z)=\mathsf{L}%
^{m}(z)^{-1}. \label{4.6}%
\end{equation}
Then%
\begin{equation}%
{\hat{\bm{E}}}%
(\mathbf{x},z)=\mathsf{R}^{e}(z)\cdot\mathbf{g}^{e}(\mathbf{x},z),\;%
{\hat{\bm{H}}}%
(\mathbf{x},z)=\mathsf{R}^{m}(z)\cdot\mathbf{g}^{m}(\mathbf{x},z). \label{4.7}%
\end{equation}
Note that%
\begin{equation}
\partial_{\mathbf{x}}\cdot\mathsf{L}^{e}(z)\cdot%
{\hat{\bm{E}}}%
(\mathbf{x},z)=z^{2}\partial_{\mathbf{x}}\cdot\{%
{\bm{\varepsilon}}%
(\mathbf{x},z)\cdot%
{\hat{\bm{E}}}%
(\mathbf{x},z)\}=z^{2}\partial_{\mathbf{x}}\cdot%
{\hat{\bm{D}}}%
(\mathbf{x},z)=iz\partial_{\mathbf{x}}\cdot%
{\bm{E}}%
(\mathbf{x},0)=0, \label{4.8}%
\end{equation}
as it should be.\newline We can make the identification (see the Appendix)%
\begin{equation}
\mathsf{P}_{1}\mathsf{R}(z)\mathsf{P}_{1}=z\mathsf{R}^{e}(z)\mathsf{P}_{1}.
\label{4.9}%
\end{equation}
Since the left hand side is a bounded operator, it follows that $\mathsf{R}%
^{e}(z)$, $\operatorname{Im}z>0$, has a closed densely defined extension,
which is in fact bounded and the same is true for \medskip$\mathsf{R}^{m}(z)$.
Note that, since $\overline{%
{\bm{\varepsilon}}%
(\mathbf{x},z)}=%
{\bm{\varepsilon}}%
(\mathbf{x},-\bar{z})$ and $\overline{%
\bm{\mu}%
(\mathbf{x},z)}=%
\bm{\mu}%
(\mathbf{x},-\bar{z})$, $\mathsf{R}^{e}(z)^{\ast}=\mathsf{R}^{e}(-\bar{z})$.
Next we introduce the Green's functions
\begin{equation}
\mathsf{G}^{e,m}(\mathbf{x,y},z)=\langle\mathbf{x}|\mathsf{R}^{e,m}%
(z)|\mathbf{y}\rangle. \label{4.10}%
\end{equation}
They are square integrable in $\mathbf{x}$ and $\mathbf{y}$, respectively, are
analytic in the open upper half plane and have the following further
properties%
\begin{equation}
\mathsf{G}(\mathbf{x,y},z)=\mathsf{G}(\mathbf{y,x},-\bar{z})^{\dag},\;\frac
{1}{2\pi}\int_{\Gamma}dz\mathsf{G}(\mathbf{x,y},z)=0,\;\frac{1}{2\pi i}%
\int_{\Gamma}dzz\mathsf{G}(\mathbf{x,y},z)=-\delta(\mathbf{x-y})\mathsf{U}.
\label{4.11}%
\end{equation}
Now
\begin{equation}
\mathsf{L}^{e,m}\cdot\mathsf{G}^{e,m}(\mathbf{x,y},z)=\delta(\mathbf{x-y}%
)\mathsf{U}. \label{4.12}%
\end{equation}
Note that, although $\partial_{\mathbf{x}}\cdot\mathsf{L}^{e}\cdot%
{\hat{\bm{E}}}%
(\mathbf{x},z)=0$, this is a special case since%
\begin{equation}
\partial_{\mathbf{x}}\cdot\mathsf{L}_{\mathbf{x}}^{e}\cdot\int d\mathbf{y}%
\mathsf{G}(\mathbf{x,y},z)\cdot\mathbf{h(y})=\partial_{\mathbf{x}}%
\cdot\mathbf{h(x}), \label{4.13}%
\end{equation}
which need not vanish for general $\mathbf{h}$.\newline

The spatially piecewise constant situation is the case where $\mathbb{R}%
^{3}=\{\cup_{j}\mathcal{M}_{j}\}\cup\{%
\text{interfaces}%
\},$ with the $\mathcal{M}_{j}$'s disjoint open sets separated by sufficiently
regular interfaces (so boundary conditions can be imposed), whereas the
susceptibilities are constant over $\mathcal{M}_{j}$,%
\begin{equation}%
{\bm{\varepsilon}}%
(\mathbf{x},z)=%
{\bm{\varepsilon}}%
_{j}(z),\;%
\bm{\mu}%
(\mathbf{x},z)=%
\bm{\mu}%
_{j}(z),\;\mathbf{x}\in\mathcal{M}_{j}. \label{4.14}%
\end{equation}

\section{Layered systems}

\subsection{General}

In the sequel we only consider the electric Green's function and we drop the
superscript $e$. We also assume that the system is isotropic (in the
anisotropic case the reduction to an expression featuring the scalar Green's
functions for $s$ and $p$ polarization is not possible in general). Here we
consider the situation that the $\mathcal{M}_{j}$'s are a number of layers
parallel to the $X_{1}X_{2}$-plane. Then the permeabilities only depend on
$x_{3}$,
\begin{equation}
\varepsilon(\mathbf{x},z)=\varepsilon(x_{3},z),\;\mu(\mathbf{x},z)=\mu
(x_{3},z). \label{5.1}%
\end{equation}
We exploit the translational invariance in the $X_{1}$ and $X_{2}$-directions.
Let $\mathbf{k}=(k_{1},k_{2},k_{3})$ and $%
{\bm{\kappa}}%
=\kappa\mathbf{e}_{%
{\bm{\kappa}}%
}=\mathbf{k}^{\perp}=(k_{1},k_{2},0)\perp\mathbf{e}_{3}$, \textbf{ }%
\begin{align}
\mathsf{G}(\mathbf{x,y},z)  &  =(2\pi)^{-2}\int d%
{\bm{\kappa}}%
\exp[-i%
{\bm{\kappa}}%
\mathbf{\cdot}(\mathbf{x^{\perp}-y^{\perp}})]\mathsf{G}_{%
{\bm{\kappa}}%
}(x_{3},y_{3},z),\nonumber\\
\mathsf{G}_{%
{\bm{\kappa}}%
}(x_{3},y_{3},z)  &  =(2\pi)^{-2}\int d\mathbf{x}^{\perp}\exp[i%
{\bm{\kappa}}%
\mathbf{\cdot(x}^{\perp}-\mathbf{y}^{\perp})]\mathsf{G}(\mathbf{x,y}%
,z),\nonumber\\
\mathbf{m}_{%
{\bm{\kappa}}%
}(x_{3},z)  &  =\int d\mathbf{x}^{\perp}\exp[i%
{\bm{\kappa}}%
\mathbf{\cdot x}^{\perp}]\mathbf{m}(\mathbf{x},z), \label{5.2}%
\end{align}
where $\mathbf{m}$ can be $%
{\bm{E}}%
$, $%
{\bm{B}}%
$, $\mathbf{g}^{e,m}$, etc. Then%
\begin{align}%
{\bm{E}}%
(\mathbf{x},t)  &  =(2\pi)^{-1}\int_{\Gamma}dz\exp[-izt]%
{\hat{\bm{E}}}%
(\mathbf{x},z)\nonumber\\
&  =(2\pi)^{-3}\int_{\Gamma}dz\exp[-izt]\int d%
{\bm{\kappa}}%
\exp[-i%
{\bm{\kappa}}%
\mathbf{\cdot}(\mathbf{x^{\perp}-y^{\perp}})]\int dy_{3}\mathsf{G}_{%
{\bm{\kappa}}%
}(x_{3},y_{3},z)\cdot\mathbf{g}_{%
{\bm{\kappa}}%
}(y_{3},z). \label{5.3}%
\end{align}
Omitting the subscript $3$ in $x_{3}$, etc., from now on, $%
{\hat{\bm{E}}}%
_{%
{\bm{\kappa}}%
}(x,z)\in\mathcal{H}=L^{2}(\mathbb{R},dx;\mathbb{C}^{3})$,%

\begin{align}
\mathbf{g}_{%
{\bm{\kappa}}%
}(y,z)  &  =iz%
{\bm{E}}%
_{%
{\bm{\kappa}}%
}(y,0)-(i%
{\bm{\kappa}}%
+\partial_{y}\mathbf{e}_{3})\times\dfrac{1}{\mu(y,z)}%
{\bm{B}}%
_{%
{\bm{\kappa}}%
}(y,0),\nonumber\\
\mathsf{L}_{%
{\bm{\kappa}}%
}\cdot%
{\hat{\bm{E}}}%
_{%
{\bm{\kappa}}%
}(x,z)  &  =\mathbf{g}_{%
{\bm{\kappa}}%
}(x,z), \label{5.4}%
\end{align}
and%
\begin{equation}
\mathsf{G}_{%
{\bm{\kappa}}%
}(x,y,z)=\langle x|\mathsf{L}_{%
{\bm{\kappa}}%
}^{-1}|y\rangle,\;\mathsf{L}_{%
{\bm{\kappa}}%
}\cdot\mathsf{G}_{%
{\bm{\kappa}}%
}(x,y,z)=\delta(x-y)\mathsf{U}. \label{5.5}%
\end{equation}
$\mathsf{L}_{%
{\bm{\kappa}}%
}$ is obtained from $\mathsf{L}^{e}$ by replacing $\mathbf{p}$ by $%
{\bm{\kappa}}%
+p\mathbf{e}_{3}=\kappa\mathbf{e}_{%
{\bm{\kappa}}%
}+p\mathbf{e}_{3}$, $p=-i\partial_{x}$. Denoting%
\begin{equation}
\zeta(x,\kappa,z)^{2}=z^{2}\varepsilon(x,z)\mu(x,z)-\kappa^{2}, \label{5.6}%
\end{equation}
we obtain%
\begin{align}
\mathsf{L}_{%
{\bm{\kappa}}%
}  &  =\{\frac{\zeta(x,\kappa,z)^{2}}{\mu(\mathbf{x},z)}-p\dfrac{1}%
{\mu(\mathbf{x},z)}p\}\mathbf{e}_{3}\times\mathbf{e}_{%
{\bm{\kappa}}%
}\mathbf{e}_{3}\times\mathbf{e}_{%
{\bm{\kappa}}%
}\nonumber\\
&  +\{z^{2}\varepsilon(\mathbf{x},z)-p\dfrac{1}{\mu(\mathbf{x},z)}%
p\}\mathbf{e}_{%
{\bm{\kappa}}%
}\mathbf{e}_{%
{\bm{\kappa}}%
}+\frac{\zeta(x,\kappa,z)^{2}}{\mu(\mathbf{x},z)}\mathbf{e}_{3}\mathbf{e}%
_{3}+p\frac{\kappa}{\mu(\mathbf{x},z)}\mathbf{e}_{%
{\bm{\kappa}}%
}\mathbf{e}_{3}+\frac{\kappa}{\mu(\mathbf{x},z)}p\mathbf{e}_{3}\mathbf{e}_{%
{\bm{\kappa}}%
}\nonumber\\
&  =\mathsf{L}_{%
{\bm{\kappa}}%
}^{s}+\mathsf{L}_{%
{\bm{\kappa}}%
}^{p}, \label{5.7}%
\end{align}
where $\mathsf{L}_{%
{\bm{\kappa}}%
}^{s}$ is the $s$-polarization part (the term with $\mathbf{e}_{3}%
\times\mathbf{e}_{%
{\bm{\kappa}}%
}\mathbf{e}_{3}\times\mathbf{e}_{%
{\bm{\kappa}}%
}$) and $\mathsf{L}_{%
{\bm{\kappa}}%
}^{p}$, the $p$-polarization part, the remainder. The corresponding
decomposition for $\mathsf{G}_{%
{\bm{\kappa}}%
}$ is%
\begin{equation}
\mathsf{G}_{%
{\bm{\kappa}}%
}(x,y,z)=\mathsf{G}_{%
{\bm{\kappa}}%
}^{s}(x,y,z)+\mathsf{G}_{%
{\bm{\kappa}}%
}^{p}(x,y,z)=\langle x|(\mathsf{L}_{%
{\bm{\kappa}}%
}^{s})^{-1}|y\rangle+\langle x|(\mathsf{L}_{%
{\bm{\kappa}}%
}^{p})^{-1}|y\rangle. \label{5.8}%
\end{equation}
It is customary to consider the scalar Green's functions associated with \ the
electric and magnetic $s$-polarization parts. However, the latter is
transverse, whereas $\mathsf{G}_{%
{\bm{\kappa}}%
}^{p}$ also contains a longitudinal component. In addition, in obtaining
atomic radiative decay rates, the full tensorial expression for the Green's
function is required and there is no simple relation between (the transverse
part of) $\mathsf{G}_{%
{\bm{\kappa}}%
}^{p}$ and the magnetic $s$-polarized Green's function. Thus we calculated
$\mathsf{G}_{%
{\bm{\kappa}}%
}^{p}$ in the Appendix with the result%
\begin{align}
\mathsf{G}_{%
{\bm{\kappa}}%
}(x\mathbf{,}y,z)  &  =\mathsf{G}_{s}(x,y,z,\kappa)+\mathsf{G}_{p}%
(x,y,z,\kappa),\nonumber\\
\mathsf{G}_{s}(x,y,z,\kappa)  &  =G_{s}(x,y,z,\kappa)\mathbf{e}_{3}%
\times\mathbf{e}_{%
{\bm{\kappa}}%
}\mathbf{e}_{3}\times\mathbf{e}_{%
{\bm{\kappa}}%
},\nonumber\\
\mathsf{G}_{p}(x,y,z,\kappa)  &  =(\mathbf{e}_{%
{\bm{\kappa}}%
}+\frac{i\kappa}{\zeta(x)^{2}}\partial_{x}\mathbf{e}_{3})(\mathbf{e}_{%
{\bm{\kappa}}%
}-\frac{i\kappa}{\zeta(y)^{2}}\partial_{y}\mathbf{e}_{3})G_{p}(x,y,z,\kappa),
\label{5.9}%
\end{align}
where $G_{s}$ and $G_{p}$ satisfy
\begin{equation}
\{z^{2}\varepsilon(x,z)-p\frac{z^{2}\varepsilon(x,z)}{\zeta(x,\kappa,z)^{2}%
}p\}G_{p}(x,y,z,\kappa)=\delta(x-y),\;\{\frac{\zeta(x,\kappa,z)^{2}}{\mu
(x,z)}-p\dfrac{1}{\mu(x,z)}p\}G_{s}(x\mathbf{,}y,z,\kappa)=\delta(x-y).
\label{5.10}%
\end{equation}
In order to obtain $G_{p}$ and $G_{s}$ we have to supplement these
differential equations with the boundary conditions at an interface. Since
$\partial_{x}(\varepsilon/\zeta^{2})\partial_{x}G_{p}$ must make sense,
$\partial_{x}G_{p}$ must exist, so we can choose $G_{p}$ to be continuous in
$x$. In addition $(\varepsilon/\zeta^{2})\partial_{x}G_{p}$ must be
differentiable, so it must also be continuous in $x$. Similarly we find that
$G_{s}$ must be continuous in $x$, as well as $\mu^{-1}\partial_{x}G_{s}$ and
again the same applies with $x$ and $y$ interchanged. These boundary
conditions can be shown to correspond to the usual boundary conditions for $%
{\bm{D}}%
$ and $%
{\bm{E}}%
$. In addition, in view of the square integrability in $x$ and $y$, there are
no exponentially increasing contributions for layers that extend to
$x=\pm\infty$.

\subsection{Two half spaces filled with different materials}

We consider the situation where the half spaces $x>0$ and $x<0$ are filled
with media characterized according to%
\begin{equation}
\varepsilon(x,z)=\left\{
\begin{array}
[c]{ll}%
\varepsilon_{+}(z), & x>0,\\
\varepsilon_{-}(z), & x<0,
\end{array}
\right.  ,\;\mu(x,z)=\left\{
\begin{array}
[c]{ll}%
\mu_{+}(z), & x>0,\\
\mu_{-}(z), & x<0.
\end{array}
\right.  \label{5.11}%
\end{equation}
We denote%
\begin{align}
\zeta_{+}(\kappa,z)^{2}  &  =z^{2}\varepsilon_{+}(z)\mu_{+}(z)-\kappa
^{2},\;\zeta_{-}(\kappa,z)^{2}=z^{2}\varepsilon_{-}(z)\mu_{-}(z)-\kappa
^{2},\nonumber\\
K_{\pm}(\kappa,z)  &  =\frac{\zeta_{\pm}(\kappa,z)}{2iz^{2}\varepsilon_{\pm
}(z)},\;L_{\pm}(\kappa,z)=\frac{\mu_{\pm}(z)}{2i\zeta_{\pm}(\kappa,z)},
\label{5.12}%
\end{align}
and introduce the Fresnel reflection coefficients%
\begin{equation}
r_{p}=\frac{\varepsilon_{-}\zeta_{+}-\varepsilon_{+}\zeta_{-}}{\varepsilon
_{-}\zeta_{+}+\varepsilon_{+}\zeta_{-}},\;r_{s}=\frac{\mu_{-}\zeta_{+}-\mu
_{+}\zeta_{-}}{\mu_{-}\zeta_{+}+\mu_{+}\zeta_{-}}. \label{5.13}%
\end{equation}
Using square integrability in $x$ and $y$ and the boundary conditions on the
interface, we obtain in the usual way%
\begin{align}
G_{p}(x,y,z)  &  =K_{+}\{\exp[i\zeta_{+}|x-y|]-r_{p}\exp[i\zeta_{+}%
(x+y)]\}\theta(x)\theta(y)+\frac{1}{iz^{2}}\frac{\zeta_{+}\zeta_{-}%
}{\varepsilon_{+}\zeta_{-}+\varepsilon_{-}\zeta_{+}}\{\exp[i\zeta_{+}%
x-i\zeta_{-}y]\theta(x)\theta(-y)\nonumber\\
&  +\exp[-i\zeta_{-}x+i\zeta_{+}y]\theta(-x)\theta(y)\}+K_{-}\{\exp[i\zeta
_{-}|x-y|]+r_{p}\exp[-i\zeta_{-}(x+y)]\}\theta(-x)\theta(-y), \label{5.14}%
\end{align}
and%
\begin{align}
G_{s}(x,y,z)  &  =L_{+}\{\exp[i\zeta_{+}|x-y|]+r_{s}\exp[i\zeta_{+}%
(x+y)]\}\theta(x)\theta(y)-i\frac{\mu_{+}\mu_{-}}{\mu_{+}\zeta_{-}+\mu
_{-}\zeta_{+}}\{\exp[i\zeta_{+}x-i\zeta_{-}y]\theta(x)\theta(-y)\nonumber\\
&  +\exp[-i\zeta_{-}x+i\zeta_{+}y]\theta(-x)\theta(y)\}+L_{-}\{\exp[i\zeta
_{-}|x-y|]-r_{s}\exp[-i\zeta_{-}(x+y)]\}\theta(-x)\theta(-y), \label{5.15}%
\end{align}
from which the $x$ and $y$ derivatives of $G_{p}(x,y,z)$, etc., present in
$\mathsf{G}_{p}(x,y,z,\kappa)$, can be obtained in explicit form.

The case where the region $x>0$ consists of vacuum and the initial state is
contained in this region is of particular interest. It applies to the
situation where an electromagnetic wavepacket in vacuum is travelling towards
the medium. Then, labeling vacuum quantities with the subscript $0$ and
deleting the subscript $-$ for quantities associated with the medium,
\begin{align}
K_{+}(\kappa,z)  &  =K_{0}(\kappa,z)=\frac{\zeta_{0}(\kappa,z)}{2iz^{2}%
},\;L_{+}(\kappa,z)=L_{0}(\kappa,z)=\frac{1}{2i\zeta_{0}(\kappa,z)}%
,\;\nonumber\\
\zeta_{0}(\kappa,z)  &  =\sqrt{z^{2}-\kappa^{2}},\;r_{p0}=\frac{\varepsilon
\zeta_{0}-\zeta}{\varepsilon\zeta_{0}+\zeta},\;r_{s0}=\frac{\mu\zeta_{0}%
-\zeta}{\mu\zeta_{0}+\zeta}, \label{5.16}%
\end{align}
so, for $y>0$,%
\begin{align}
G_{p}(x,y,z)  &  =K_{0}\left\{  \exp[i\zeta_{0}|x-y|]-r_{p0}\exp[i\zeta
_{0}(x+y)]\right\}  \theta(x)+\frac{1}{iz^{2}}\frac{\zeta_{0}\zeta}%
{\zeta+\varepsilon\zeta_{0}}\exp[-i\zeta x+i\zeta_{0}y]\theta(-x),\nonumber\\
G_{s}(x,y,z)  &  =L_{0}\{\exp[i\zeta_{0}|x-y|]+r_{s0}\exp[i\zeta
_{0}(x+y)]\}\theta(x)-i\frac{\mu}{\zeta+\mu\zeta_{0}}\exp[-i\zeta x+i\zeta
_{0}y]\theta(-x). \label{5.17}%
\end{align}

\section{The NIM situation}

We continue our investigation of the half space case, assuming that
\begin{equation}
\varepsilon(z)=\mu(z)=1-\frac{\Omega^{2}}{z^{2}-\omega_{0}^{2}}, \label{6.1}%
\end{equation}
the dispersive Lorentz case, and rewrite%
\begin{equation}
\Phi(z)=\frac{1}{\zeta(\kappa,z)+\varepsilon(z)\zeta_{0}(\kappa,z)}%
=\frac{\zeta-\varepsilon\zeta_{0}}{\zeta^{2}-\varepsilon^{2}\zeta_{0}^{2}%
}=\frac{\zeta-\varepsilon\zeta_{0}}{(z^{2}\varepsilon^{2}-\kappa
^{2})-\varepsilon^{2}(z^{2}-\kappa^{2})}=\frac{\zeta-\varepsilon\zeta_{0}%
}{\kappa^{2}}\frac{1}{\varepsilon-1}\frac{1}{\varepsilon+1}. \label{6.2}%
\end{equation}
Note that this expression can become infinite if $\varepsilon^{2}=1$. In case
$\varepsilon=1$ we are back to the vacuum case and $\zeta-\varepsilon\zeta
_{0}=0$. But if $\varepsilon=\mu=-1$, the NIM case,
\begin{equation}
z=\hat{\omega}_{\pm}=\pm\hat{\omega},\;\hat{\omega}=\sqrt{\omega_{0}^{2}%
+\frac{1}{2}\Omega^{2}}, \label{6.3}%
\end{equation}
and $\Phi(z)$ can become infinite. Now%
\begin{equation}
\frac{1}{\varepsilon(z)+1}=\frac{1}{2}\frac{z^{2}-\omega_{0}^{2}}%
{(z-\hat{\omega})(z+\hat{\omega})},\;\Phi(z)=\frac{\zeta-\varepsilon\zeta_{0}%
}{(z-\hat{\omega})(z+\hat{\omega})}\frac{z^{2}-\omega_{0}^{2}}{2\kappa
^{2}\{\varepsilon(z)-1\}}, \label{6.4}%
\end{equation}
so we encounter poles in $z=\pm\hat{\omega}$. Next we study the behavior of
$\Phi(z)=\Phi(\omega+i\delta)$ as $\delta\downarrow0$. We start with
$\zeta(z)$. In order to obtain the square root in the limit $\delta
\downarrow0$ we must know the signs of $\zeta(\omega)^{2}$ and $\partial
_{\omega}\zeta(\omega)^{2}$ in%
\begin{equation}
\zeta(\omega+i\delta)^{2}=\zeta(\omega)^{2}+i\delta\partial_{\omega}%
\zeta(\omega)^{2}+\mathcal{O}(\delta^{2}), \label{6.5}%
\end{equation}
where%
\begin{equation}
\zeta(z)^{2}=z^{2}\left(  1-\frac{\Omega^{2}}{z^{2}-\omega_{0}^{2}}\right)
^{2}-\kappa^{2},\;\partial_{z}\zeta(z)^{2}=2z\left(  1-\frac{\Omega^{2}}%
{z^{2}-\omega_{0}^{2}}\right)  \left(  1+\frac{\Omega^{2}}{z^{2}-\omega
_{0}^{2}}+\frac{2\omega_{0}^{2}\Omega^{2}}{(z^{2}-\omega_{0}^{2})^{2}}\right)
. \label{6.6}%
\end{equation}
Since both quantities have definite parity it suffices to consider the case
$\omega\geqslant0$. We note that $\zeta(0)^{2}=-\kappa^{2}$ and that
$\zeta(\omega)^{2}$ increases to $+\infty$ as $\omega$ approaches $\omega_{0}%
$. Then it decreases again to reach the value $-\kappa^{2}$ for $\omega
=\tilde{\omega}=\sqrt{\omega_{0}^{2}+\Omega^{2}}$. Beyond this value it
increases again to tend to $+\infty$ as $\omega\rightarrow+\infty$. Thus
$\zeta(\omega)^{2}$ has three zero's, $\omega_{a}\in(0,\omega_{0})$,
$\omega_{b}\in(\omega_{0},\tilde{\omega})$ and $\omega_{c}>\tilde{\omega}$.
Since $\zeta(\omega)^{2}=0$ corresponds to a third order equation in
$\omega^{2}$ these are the full set of zero's. $\partial_{\omega}\zeta
(\omega)^{2}$ vanishes in $\omega=0$, then tends to $+\infty$ as $\omega$
reaches $\omega_{0}$, where it switches sign and increases to $0$ in
$\tilde{\omega}$, whereupon it remains positive and eventually tends to
$+\infty$. Denoting%
\begin{equation}
\rho(\omega)=\sqrt{\left\vert \omega^{2}\left(  1-\frac{\Omega^{2}}{\omega
^{2}-\omega_{0}^{2}}\right)  ^{2}-\kappa^{2}\right\vert }, \label{6.7}%
\end{equation}
we obtain ($+$ indicates that a quantity is positive, $-$ that it is negative)%
\begin{align}
&
\begin{tabular}
[c]{|l|l|l|l|l|l|l|}\hline
$\omega\in$ & $>+\omega_{c}$ & $(+\tilde{\omega},+\omega_{c})$ & $(+\omega
_{b},+\tilde{\omega})$ & $(+\omega_{0},+\omega_{b})$ & $(+\omega_{a}%
,+\omega_{0})$ & $(0,+\omega_{a})$\\\hline
$\zeta(\kappa,\omega)^{2}$ & $+$ & $-$ & $-$ & $+$ & $+$ & $-$\\\hline
$\operatorname{Im}\zeta(\kappa,\omega+i\delta)^{2}$ & $+$ & $+$ & $-$ & $-$ &
$+$ & $+$\\\hline
$\zeta(\kappa,\omega)$ & $\rho(\omega)$ & $i\rho(\omega)$ & $i\rho(\omega)$ &
$-\rho(\omega)$ & $\rho(\omega)$ & $i\rho(\omega)$\\\hline
\end{tabular}
\ \ ,\ \nonumber\\
&
\begin{tabular}
[c]{|l|l|l|l|l|l|l|}\hline
$\omega\in$ & $<-\omega_{c}$ & $(-\omega_{c},-\tilde{\omega})$ &
$(-\tilde{\omega},-\omega_{b})$ & $(-\omega_{b},-\omega_{0})$ & $(-\omega
_{0},-\omega_{a})$ & $(-\omega_{a},0)$\\\hline
$\zeta(\kappa,\omega)^{2}$ & $+$ & $-$ & $-$ & $+$ & $+$ & $-$\\\hline
$\operatorname{Im}\zeta(\kappa,\omega+i\delta)^{2}$ & $-$ & $-$ & $+$ & $+$ &
$-$ & $-$\\\hline
$\zeta(\kappa,\omega)$ & $-\rho(\omega)$ & $i\rho(\omega)$ & $i\rho(\omega)$ &
$\rho(\omega)$ & $-\rho(\omega)$ & $i\rho(\omega)$\\\hline
\end{tabular}
\ \ .\ \label{6.8}%
\end{align}
In comparison, with $\rho_{0}(\omega)=\sqrt{|\omega^{2}-\kappa^{2}|}$, so
$\rho(\hat{\omega})=\rho_{0}(\hat{\omega})$,%
\begin{equation}%
\begin{tabular}
[c]{|l|l|l|l|l|}\hline
$\omega\in$ & $>\kappa$ & $(0,+\kappa)$ & $(-\kappa,0)$ & $<-\kappa$\\\hline
$\zeta_{0}(\kappa,\omega)$ & $\rho_{0}(\omega)$ & $i\rho_{0}(\omega)$ &
$i\rho_{0}(\omega)$ & $-\rho_{0}(\omega)$\\\hline
\end{tabular}
\ \ \ \ \ \ \ \ \ \ \ \ \ \ \ . \label{6.9}%
\end{equation}
\newline Since $\hat{\omega}=\sqrt{\omega_{0}^{2}+\frac{1}{2}\Omega^{2}}$ we
have $\hat{\omega}\in(\omega_{0},\tilde{\omega})$. For $\hat{\omega}>\kappa$,
$\zeta(\kappa,\hat{\omega})^{2}=\hat{\omega}^{2}-\kappa^{2}>0$ and
$\hat{\omega}\in(\omega_{0},\omega_{b})$, $\zeta(\kappa,\hat{\omega}%
)=-\rho(\hat{\omega})=-\sqrt{\hat{\omega}^{2}-\kappa^{2}}$, whereas for
$\hat{\omega}<\kappa$, $\zeta(\kappa,\hat{\omega})^{2}<0$ and $\hat{\omega}%
\in(\omega_{b},\tilde{\omega})$, $\zeta(\kappa,\hat{\omega})=i\rho(\hat
{\omega})$. Similar results follow for $-\hat{\omega}$ resulting in%
\begin{equation}%
\begin{tabular}
[c]{|l|l|l|}\hline
$\hat{\omega}\in$ & $>\kappa$ & $(0,+\kappa)$\\\hline
$\zeta(\kappa,+\hat{\omega})$ & $-\rho(\hat{\omega})$ & $i\rho(\hat{\omega}%
)$\\\hline
$\zeta(\kappa,-\hat{\omega})$ & $+\rho(\hat{\omega})$ & $i\rho(\hat{\omega}%
)$\\\hline
$\zeta_{0}(\kappa,+\hat{\omega})$ & $+\rho(\hat{\omega})$ & $i\rho(\hat
{\omega})$\\\hline
$\zeta_{0}(\kappa,-\hat{\omega})$ & $-\rho(\hat{\omega})$ & $i\rho(\hat
{\omega})$\\\hline
$\zeta_{-}(\hat{\omega})-\varepsilon(\hat{\omega})\zeta_{0}(\hat{\omega})$ &
$0$ & $2i\rho(\hat{\omega})$\\\hline
$\zeta_{-}(-\hat{\omega})-\varepsilon(-\hat{\omega})\zeta_{0}(-\hat{\omega})$
& $0$ & $2i\rho(\hat{\omega})$\\\hline
$\zeta(\hat{\omega})+\varepsilon(\hat{\omega})\zeta_{0}(\hat{\omega})$ &
$-2\rho(\hat{\omega})$ & $0$\\\hline
$\zeta(-\hat{\omega})+\varepsilon(-\hat{\omega})\zeta_{0}(-\hat{\omega})$ &
$+2\rho(\hat{\omega})$ & $0$\\\hline
\end{tabular}
\ \ .\ \ \ \ \ \ \ \label{6.10}%
\end{equation}
For $\hat{\omega}>\kappa$ and $y>0$ we obtain%
\begin{align}
G_{p}(x,y,\pm\hat{\omega})  &  =\pm\frac{\rho(\hat{\omega})}{2i\hat{\omega
}^{2}}\{\exp[\pm i\rho(\hat{\omega})|x-y|]\theta(x)+\exp[\pm i\rho(\hat
{\omega})(x+y)]\theta(-x)\}\nonumber\\
G_{s}(x,y,\pm\hat{\omega})  &  =\pm\frac{1}{2i\rho(\hat{\omega})}\{\exp[\pm
i\rho(\hat{\omega})|x-y|]\theta(x)+\exp[\pm i\rho(\hat{\omega})(x+y)]\theta
(-x)\}. \label{6.11}%
\end{align}
On the other hand, for $\hat{\omega}<\kappa$, $\Phi(z)$ becomes infinite for
$z=\pm\hat{\omega}$ since $\zeta(\kappa,z)+\varepsilon(z)\zeta_{0}(\kappa,z)$
vanishes in $\pm\hat{\omega}$. Now%
\begin{equation}
\Phi(z)=\frac{\zeta-\varepsilon\zeta_{0}}{(z-\hat{\omega})(z+\hat{\omega}%
)}\frac{z^{2}-\omega_{0}^{2}}{2\kappa^{2}\{\varepsilon(z)-1\}}\overset
{z\rightarrow\pm\hat{\omega}}{\sim}\frac{\rho(\hat{\omega})\Omega^{2}%
}{4i\kappa^{2}}\frac{1}{(z-\hat{\omega})(z+\hat{\omega})}, \label{6.12}%
\end{equation}
so we encounter poles in $\pm\hat{\omega}$. Thus
\begin{align}
&  G_{p}(x,y,z)=K_{0}\left\{  \exp[i\zeta_{0}|x-y|]+\frac{\zeta-\varepsilon
\zeta_{0}}{\zeta+\varepsilon\zeta_{0}}\exp[i\zeta_{0}(x+y)]\right\}
\theta(x)+\frac{1}{iz^{2}}\frac{\zeta_{0}\zeta}{\zeta+\varepsilon\zeta_{0}%
}\exp[-i\zeta x+i\zeta_{0}y]\theta(-x)\nonumber\\
&  =K_{0}\exp[i\zeta_{0}|x-y|]\theta(x)+\frac{1}{\zeta+\varepsilon\zeta_{0}%
}\{\kappa_{0}(\zeta-\varepsilon\zeta_{0})\exp[i\zeta_{0}(x+y)]\theta
(x)+\frac{1}{iz^{2}}\zeta_{0}\zeta\exp[-i\zeta x+i\zeta_{0}y]\theta
(-x)\}\nonumber\\
&  \sim\frac{\rho(\hat{\omega})}{2\hat{\omega}^{2}}\exp[-\rho(\hat{\omega
})|x-y|]\theta(x)+\frac{\rho(\hat{\omega})^{3}\Omega^{2}}{4\hat{\omega}%
^{2}\kappa^{2}}\frac{1}{(z-\hat{\omega})(z+\hat{\omega})}\{\exp[-\rho
(\hat{\omega})(x+y)]\theta(x)+\exp[+\rho(\hat{\omega})(x-y)]\theta
(-x)\}\nonumber\\
&  G_{s}(x,y,z)=L_{0}\left\{  \exp[i\zeta_{0}|x-y|]-\frac{\zeta-\mu\zeta_{0}%
}{\zeta+\mu\zeta_{0}}\exp[i\zeta_{0}(x+y)]\right\}  \theta(x)-i\frac{\mu
}{\zeta+\mu\zeta_{0}}\exp[-i\zeta x+i\zeta_{0}y]\theta(-x)\nonumber\\
&  =L_{0}\exp[i\zeta_{0}|x-y|]\theta(x)-\frac{1}{\zeta+\mu\zeta_{0}}%
\{\lambda_{0}(\zeta-\mu\zeta_{0})\exp[i\zeta_{0}(x+y)]\theta(x)+i\mu
\exp[-i\zeta x+i\zeta_{0}y]\theta(-x)\}\nonumber\\
&  \overset{z\rightarrow\pm\hat{\omega}}{\sim}-\frac{1}{2\rho(\hat{\omega}%
)}\exp[-\rho(\hat{\omega})|x-y|]\theta(x)+i\frac{\rho(\hat{\omega})\Omega^{2}%
}{4i\kappa^{2}}\frac{1}{(z-\hat{\omega})(z+\hat{\omega})}\{\exp[-\rho
(\hat{\omega})(x+y)]\theta(x)+\exp[+\rho(\hat{\omega})(x-y)]\theta(-x)\}.
\label{6.13}%
\end{align}
Hence, in the reflection case $x,y>0$, and for $\hat{\omega}>\kappa$,
\begin{equation}
G_{p}(x,y,\pm\hat{\omega})=\pm\frac{\rho(\hat{\omega})}{2iz^{2}}\{\exp[\pm
i\rho(\hat{\omega})|x-y|],\;G_{s}(x,y,\pm\hat{\omega})=\pm\frac{1}{2i\rho
(\hat{\omega})}\{\exp[\pm i\rho(\hat{\omega})|x-y|], \label{6.14}%
\end{equation}
where the term responsible for reflection is absent, i.e., \textit{there is no
reflection at the frequencies} $\pm\hat{\omega}$ \textit{for which}
$\varepsilon(z)=\mu(z)=-1$. \newline On the other hand, for $\hat{\omega
}<\kappa$, $x,y>0$,%
\begin{align}
&  G_{p}(x,y,z)\overset{z\rightarrow\pm\hat{\omega}}{\sim}\frac{\rho
(\hat{\omega})}{2\hat{\omega}^{2}}\exp[-\rho(\hat{\omega})|x-y|]+\frac
{\rho(\hat{\omega})^{3}\Omega^{2}}{4\hat{\omega}^{2}\kappa^{2}}\frac
{1}{(z-\hat{\omega})(z+\hat{\omega})}\exp[-\rho(\hat{\omega}%
)(x+y)],\nonumber\\
&  G_{s}(x,y,z)\overset{z\rightarrow\pm\hat{\omega}}{\sim}-\frac{1}{2\rho
(\hat{\omega})}\exp[-\rho(\hat{\omega})|x-y|]+\frac{\rho(\hat{\omega}%
)\Omega^{2}}{4\kappa^{2}}\frac{1}{(z-\hat{\omega})(z+\hat{\omega})}\exp
[-\rho(\hat{\omega})(x+y)], \label{6.15}%
\end{align}
so now the reflection term is still present but we encounter the damped
behavior, typical for the evanescent situation. \newline Next we consider
refraction (transmission into the lower half space). Here $y>0>x$, and for
\ $\hat{\omega}>\kappa$,%
\begin{equation}
G_{p}(x,y,\pm\hat{\omega})=\pm\frac{\rho(\hat{\omega})}{2i\hat{\omega}^{2}%
}\exp[\pm i\rho(\hat{\omega})(x+y)],\;G_{s}(x,y,\pm\hat{\omega})=\pm\frac
{1}{2i\rho(\hat{\omega})}\exp[\pm i\rho(\hat{\omega})(x+y)]. \label{6.16}%
\end{equation}
Now
\begin{equation}
\frac{i\kappa}{\zeta(\hat{\omega})^{2}}\partial_{x}G_{p}(x\mathbf{,}%
y,z)=\mp\frac{\kappa}{\rho(\hat{\omega})}, \label{6.17}%
\end{equation}
leading to%
\begin{equation}
\mathsf{G}_{%
{\bm{\kappa}}%
}(x\mathbf{,}y,z)\overset{z\rightarrow\pm\hat{\omega}}{\rightarrow}\pm\frac
{1}{2i\rho(\hat{\omega})}\exp[\pm i\rho(\hat{\omega})(y+x)]\{\frac{\rho
(\hat{\omega})^{2}}{\hat{\omega}^{2}}(\mathbf{e}_{%
{\bm{\kappa}}%
}\mp\frac{\kappa}{\rho(\hat{\omega})}\mathbf{e}_{3})(\mathbf{e}_{%
{\bm{\kappa}}%
}\mp\frac{\kappa}{\rho(\hat{\omega})}\mathbf{e}_{3})+\mathbf{e}_{3}%
\times\mathbf{e}_{%
{\bm{\kappa}}%
}\mathbf{e}_{3}\times\mathbf{e}_{%
{\bm{\kappa}}%
}\}. \label{6.18}%
\end{equation}
whereas in the vacuum case%
\begin{equation}
\mathsf{G}_{%
{\bm{\kappa}}%
}(x\mathbf{,}y,\pm\hat{\omega})=\pm\frac{1}{2i\rho(\hat{\omega})}\exp[\pm
i\rho(\hat{\omega})(y-x)]\{\frac{\rho(\hat{\omega})^{2}}{\hat{\omega}^{2}%
}(\mathbf{e}_{%
{\bm{\kappa}}%
}\pm\frac{\kappa}{\rho(\hat{\omega})}\mathbf{e}_{3})(\mathbf{e}_{%
{\bm{\kappa}}%
}\pm\frac{\kappa}{\rho(\hat{\omega})}\mathbf{e}_{3})+\mathbf{e}_{3}%
\times\mathbf{e}_{%
{\bm{\kappa}}%
}\mathbf{e}_{3}\times\mathbf{e}_{%
{\bm{\kappa}}%
}\}. \label{6.19}%
\end{equation}
Comparing the two we note that $x$ has changed to $-x$ and $\kappa/\rho
(\hat{\omega})$ to $-\kappa/\rho(\hat{\omega})$, showing the anomalous
behavior found earlier for NIM systems. This behavior becomes more direct in a
scattering formalism where it would show up in the corresponding scattering
amplitude. However, setting up a scattering formalism, although a
straightforward matter using the auxiliary field approach (the dielectric case
was treated earlier in \cite{LAD}), involves a substantial amount of
bookkeeping. This is mainly due to the existence of two different scattering
channels for reflection and transmission. The scattering situation is somewhat
less complicated for a single NIM layer, where the transmitted wave eventually
is moving in vacuum again.

If \ $\hat{\omega}<\kappa$, for $y>0>x$,
\begin{equation}
G_{p}(x,y,z)\overset{z\rightarrow\pm\hat{\omega}}{\sim}\frac{\rho(\hat{\omega
})^{3}\Omega^{2}}{4\hat{\omega}^{2}\kappa^{2}(z^{2}-\hat{\omega}^{2})}%
\exp[-\rho(\hat{\omega})(y-x)],\;G_{s}(x,y,z)\overset{z\rightarrow\pm
\hat{\omega}}{\sim}\frac{\rho(\hat{\omega})\Omega^{2}}{4\kappa^{2}(z^{2}%
-\hat{\omega}^{2})}\exp[-\rho(\hat{\omega})(y-x)], \label{6.20}%
\end{equation}
once more showing evanescent behavior. In retrieving $%
{\bm{E}}%
(\mathbf{x},t)$, the pole contributions in the Green's function give rise to
terms oscillating in time according to $\exp[\pm i\hat{\omega}t]$, so no
damping occurs in the time dependence, a property observed earlier by Pendry
\cite{Pendry} for the case of a single layer.\newline

\section{Discussion}

\subsection{Summary of results}

We started off with a system characterized by general causal, passive,
susceptibilities $\chi_{e}(\mathbf{x},t)$ and $\chi_{m}(\mathbf{x},t)$ and
showed, using the auxiliary field approach, that $%
{\bm{E}}%
(\mathbf{x},t)$ and $%
{\bm{H}}%
(\mathbf{x},t)$ have a proper time evolution. If they are square integrable at
the initial time this remains true at all later times so possible
singularities are square integrable and the electromagnetic energy remains
finite. We then specialized to layered systems using a Laplace transformed
formalism. We expressed the Green's function, which is a tensor, in terms of
the two scalar functions $G_{s}(x,y,z)$ and $G_{p}(x,y,z)$. A consequence of
the auxiliary field setup is that the Helmholtz Green's function is square
integrable in both coordinates. In particular this is true in the evanescent
case. \newline We then studied the special situation where one half space is
vacuum and the other filled with a medium. Restricting ourselves to scalar
permeabilities given by a single dispersive Lorentz term,%
\begin{equation}
\varepsilon(z)=\mu(z)=1-\frac{\Omega^{2}}{z^{2}-\omega_{0}^{2}}%
,\;\operatorname{Im}z>0, \label{7.1}%
\end{equation}
which take on the value $-1$ for $z=\pm\hat{\omega}=\pm(\omega_{0}^{2}%
+\frac{1}{2}\Omega^{2})^{1/2}$, we then evaluated $G_{s}(x,y,z)$ and
$G_{p}(x,y,z)$ as $z$ approaches these values and found, for the
non-evanescent case, a typical NIM behavior, the reflected field vanishes
whereas the transmitted field behaves anomalously, in accordance with earlier
results on NIM systems. As is well known, although $n^{2}=\varepsilon(\pm
\hat{\omega})\mu(\pm\hat{\omega})=1$, the NIM case is different from the
vacuum situation as is seen from the refractive behavior. But the absence of
reflection in the radiative regime is shared by both.

On the other hand the Green's function has poles at $\pm\hat{\omega}$ in the
evanescent situation. They lead to oscillating terms proportional to $\exp[\pm
i\hat{\omega}t]$ and the reflective part of the Green's function no longer
vanishes. There is no damping in the temporal behavior as noted earlier by
Pendry \cite{Pendry}. This is obvious since the former only occurs in
absorptive media, the spatial fall-off occurring in evanescent situations is
completely unrelated to the temporal decay found in the absorptive case.

Note that in the present setup the NIM case is a special situation occurring
for two discrete frequencies $\pm\hat{\omega}$ among a whole set where no NIM
behavior takes place. The general expression for the permeabilities of a
dispersive, non-absorptive, system for equal $\varepsilon$ and $\mu$ is a
(possibly infinite) sum of dispersive Lorentz terms,
\begin{equation}
\varepsilon(z)=\mu(z)=1-\sum_{n}\frac{\Omega_{n}^{2}}{z^{2}-\omega_{0n}^{2}},
\label{7.2}%
\end{equation}
and in this case there is a larger set of frequencies $\hat{\omega}_{n}$ for
which $\varepsilon=\mu=-1$. But between these frequencies $\varepsilon$ and
$\mu$ vary wildly, in particular there is always a $\omega_{0n}$ (for which
there is no transmission) between two subsequent $\hat{\omega}_{n}$'s. This
may spoil the idea of obtaining an interval for which this relation is
approximately valid. Indeed, if in $z=\omega+i\delta$, $\omega$ approaches
$\omega_{0n}$, then%
\begin{equation}
\varepsilon(z)\overset{\delta\downarrow0}{\sim}\frac{i\Omega_{n}^{2}}%
{2\omega_{0n}}\frac{1}{\delta}. \label{7.3}%
\end{equation}
In the half space case this results in
\begin{equation}
\frac{1}{\zeta+\varepsilon\zeta_{0}}\overset{\delta\downarrow0}{\rightarrow
}0,\;\frac{\zeta-\varepsilon\zeta_{0}}{\zeta+\varepsilon\zeta_{0}}%
\overset{\delta\downarrow0}{\rightarrow}-1, \label{7.4}%
\end{equation}
so \textsf{G}$_{%
{\bm{\kappa}}%
}(x,y,z)$ vanishes in this limit if $x<0$ and $y>0$. Thus the situation is
opposite to the NIM case. In that case there is no reflection, whereas here
the transmission vanishes (perfect reflector).

We did not consider absorption in the NIM case. This is straightforward to do
along the same lines but since the medium extends over a half space, the
transmitted field will die out. The two poles \ $\pm\hat{\omega}$\ now acquire
a negative imaginary part, so the Green's function remains finite for all
frequencies and the reflection term in the Green's function no longer
vanishes. We intend to come back to this situation for the single layer case,
where a transmitted field, although attenuated, is still present. Moreover
this case can fairly easily be treated in terms of a scattering formalism,
leading to scattering amplitudes for reflection and transmission, which should
show NIM behavior in the dispersive case for appropriate frequencies.

\subsection{Discrete eigenvalues of \textsf{K}, surface modes and radiative
atomic decay}

We found earlier that the Helmholtz Green's function of a NIM system had poles
in $z=\pm\hat{\omega}$. Usually poles in a Green's function originate from
discrete eigenvalues of the original operator and this is precisely what
happens here. In the more general case that $\lambda_{e}\neq\lambda_{m}$ and
$\Omega_{e}\neq\Omega_{m}$ the condition $\varepsilon(z)\mu(z)=1$ again gives
the solutions $z=\pm\hat{\omega}$ but now%
\begin{equation}
\hat{\omega}^{2}=\dfrac{\lambda_{e}^{2}\Omega_{m}^{2}+\lambda_{m}^{2}%
\Omega_{e}^{2}+\Omega_{e}^{2}\Omega_{m}^{2}}{\Omega_{e}^{2}+\Omega_{m}^{2}}.
\label{7.5}%
\end{equation}
It is straightforward to show that in the half space case $\pm\hat{\omega}$
are eigenvalues of \textsf{K} with associated eigenfunctions proportional to%
\begin{equation}
\exp[i%
{\bm{\kappa}}%
\cdot\mathbf{x}^{\perp}]\exp[-\sqrt{\kappa^{2}-\hat{\omega}^{2}}%
|x_{3}|],\;\kappa>\hat{\omega}, \label{7.6}%
\end{equation}
i.e., surface modes. Since $\hat{\omega}$ does not depend on $\kappa$, a
general theorem about direct integral decompositions tells us that $\pm
\hat{\omega}$ are discrete eigenvalues of \textsf{K} with infinite degeneracy.
This can also be seen directly since by superposition we can construct an
infinite orthonormal set of square integrable functions $\{\mathbf{f}%
_{n}(\mathbf{x})\}$,%
\begin{equation}
\mathbf{f}_{n}(\mathbf{x})=\int d%
{\bm{\kappa}}%
\rho_{n}(%
{\bm{\kappa}}%
)\exp[i%
{\bm{\kappa}}%
\cdot\mathbf{x}^{\perp}]\exp[-\sqrt{\kappa^{2}-\hat{\omega}^{2}}|x_{3}|].
\label{7.7}%
\end{equation}
Here the question arises as to what happens if there are whole sets of
dispersive Lorentz terms in the susceptibilities. Then there will be sets of
solutions $\{\pm\hat{\omega}_{n}\}$, but the condition $\kappa>\hat{\omega}$
needs refinement.

Excited atoms in vacuum decay by photon emission providing there is no
selection rule forbidding the transition. It is well known that the radiative
decay constant $\gamma_{vac}$ changes if we no longer have vacuum. Examples
are atoms in a cavity or near a material such as a dielectric. This is caused
by alterations in the field modes relative to the vacuum case. In simple
situations, making the dipole and isotropic approximation, $\gamma$ can be
expressed in terms of the imaginary part of the trace (\textrm{Tr}) of the
Green's function as a matrix,%
\begin{equation}
\gamma\sim\operatorname{Im}\mathrm{Tr}\mathsf{G}(\mathbf{x,x},\omega_{tr}),
\label{7.8}%
\end{equation}
where $\mathbf{x}$ is the atomic position and $\omega_{tr}$ the atomic
transition frequency. In case the permeabilities are frequency-independent,
$\operatorname{Im}\mathrm{Tr}\mathsf{G}(\mathbf{x,x},\omega)$ is the local
density of states. In \cite{LAD} the above result was obtained for
dielectrics. The same procedure, involving quantization of the fields, can be
used in the present case with the same result (the case of an atom embedded in
a magnetodielectric material was considered by the Jena group \cite{Jena}).
Here we make a few remarks about the layered case. We have%
\begin{equation}
\operatorname{Im}\mathrm{Tr}\mathsf{G}(\mathbf{x,x},\omega_{tr})=\int d%
{\bm{\kappa}}%
\operatorname{Im}\mathrm{Tr}\mathsf{G}_{%
{\bm{\kappa}}%
}(x,x,\omega_{tr}), \label{7.9}%
\end{equation}
and consider the dispersive half space situation with the atom in vacuum close
to the interface. Since the Green's function for $x,y>0$\ is the sum of a
vacuum and reflective part, we can write $\gamma=\gamma_{vac}+\gamma_{ref}$.
\ At the NIM frequency, $\omega_{tr}=\hat{\omega}$, there is no reflection in
the propagating regime so in the $%
{\bm{\kappa}}%
$-integral only $\kappa>\hat{\omega}$ can contribute. Since $\operatorname{Im}%
\mathsf{G}_{%
{\bm{\kappa}}%
}(x,x,\omega_{tr}+i0)$ becomes infinite for $\omega_{tr}=\hat{\omega}$, due to
the presence of the $\delta$-functions $\delta(\omega\pm\hat{\omega})$, we
obtain an infinite result for $\gamma_{ref}$. Such an infinite local density
of states was also encountered in \cite{Perfect}, where a so-called perfect
corner reflector was considered. However, this result is incorrect. Upon
quantization, an excited atom with $\omega_{tr}=\hat{\omega}$ can decay
radiatively but also transitions to the above bound states of \textsf{K} are
possible. The latter process is an oscillatory one and we expect decay but
modulated by oscillations. However, there are no infinities.

Another point is that the situation is highly anisotropic so the isotropy
approximation, which is used to obtain Eq. (7.8) becomes doubtful. Clearly the
radiative decay problem needs further study.

\subsection{Fixed frequency model}

In the introduction we mentioned that using a simple model with fixed, i.e.,
frequency independent $\varepsilon=\mu=-1$ can give rise to problems in
calculating the Green's function. Indeed, our formalism indicates that, in the
evanescent case, there are poles in $\pm\hat{\omega}$, so the Green's function
becomes infinite and the simple model breaks down. As we have seen, in
retrieving $%
{\bm{E}}%
(\mathbf{x},t)$ a finite result emerges and the responsible frequency
integration mechanism is absent in the model. In addition there is a second
flaw. With%
\begin{equation}
\varepsilon=\mu=\varepsilon\left(  x_{3}\right)  =\mu(x_{3})=\left\{
\begin{array}
[c]{ll}%
+1, & x_{3}>0,\\
-1, & x_{3}<0,
\end{array}
\right.  \label{7.10}%
\end{equation}
the conserved energy%
\begin{equation}
\mathcal{E}=\frac{1}{2}\int d\mathbf{x}\{\varepsilon\left(  x_{3}\right)
{\bm{E}}%
(\mathbf{x},t)^{2}+\mu(x_{3})%
{\bm{B}}%
(\mathbf{x},t)^{2}\} \label{7.11}%
\end{equation}
is no longer positive definite and we can no longer base an inner product and
associated Hilbert space formalism on this quantity. We still can use the
inner product%
\begin{equation}
(\mathbf{f,g})=\int d\mathbf{xf(x)\cdot}\overline{\mathbf{g(x)}}, \label{7.12}%
\end{equation}
but now the candidate for the generator of the time evolution, determined by
the usual boundary conditions, is no longer selfadjoint, so the existence of a
time evolution comes into question. Of course it is possible to introduce a
Krein space with inner product based on $\mathcal{E}$ but this does not solve
this problem. An alternative is to restrict \textsf{K} to the eigenspaces
associated with $\pm\hat{\omega}$, which leads to a correct time evolution.
Then we have a fixed frequency model but now the part of \textsf{K}, relevant
for radiative decay, is missing.

\subsection{Response to an external source}

We consider the time evolution due to an external source. In general the
source is given by charge and current densities $\rho_{ext}(\mathbf{x},t)$ and
$%
{\bm{J}}%
_{ext}(\mathbf{x},t)$, which are related by the conservation law%
\begin{equation}
\partial_{t}\rho_{ext}(\mathbf{x},t)+\partial_{\mathbf{x}}\cdot%
{\bm{J}}%
_{ext}(\mathbf{x},t)=0. \label{7.13}%
\end{equation}
We assume that the source quantities vanish for $t\leqslant t_{0}$, so the
same holds for the fields. Then
\begin{equation}
\partial_{t}%
{\bm{D}}%
(\mathbf{x},t)=\partial_{\mathbf{x}}\times%
{\bm{H}}%
(\mathbf{x},t)-%
{\bm{J}}%
_{ext}(\mathbf{x},t),\;\partial_{t}%
{\bm{F}}%
(t)=-i\mathsf{K}\cdot%
{\bm{F}}%
(t)-%
{\bm{G}}%
(t), \label{7.14}%
\end{equation}
where $%
{\bm{G}}%
_{1}(t)=%
{\bm{J}}%
_{ext}(\mathbf{x},t)$, whereas its other components vanish. Since $%
{\bm{F}}%
(t_{0})=0$, Duhamel's formula gives%

\begin{equation}%
{\bm{F}}%
(t)\;=\int_{t_{0}}^{t}ds\exp[-i\mathsf{K}(t-s)]\cdot%
{\bm{G}}%
(s). \label{7.15}%
\end{equation}
We are interested in the behaviour of $%
{\bm{F}}%
(t)$ for large $t$. This depends on the nature of the spectrum of $\mathsf{K}%
$. We assume that $\mathsf{K}$ does not have singular continuous spectrum, so%
\begin{equation}
\mathsf{K}=\sum_{n}\lambda_{n}\mathsf{P}_{n}+\int\lambda%
{\bm{E}}%
_{ac}(d\lambda)=\sum_{n}\lambda_{n}\mathsf{P}_{n}+\sum_{\alpha}\int
d\lambda\lambda|\mathbf{u}_{\lambda\alpha}\rangle\langle\mathbf{u}%
_{\lambda\alpha}|. \label{7.16}%
\end{equation}
At this point we set $t_{0}=0$ and assume that%
\begin{equation}%
{\bm{G}}%
(t)=f(t)%
{\bm{G}}%
_{0},\;%
{\bm{G}}%
_{0}\in\mathcal{K}, \label{7.17}%
\end{equation}
where the Fourier transform $\tilde{f}(\omega)$ in%
\begin{equation}
f(t)=\int d\omega\exp[-i\omega t]\tilde{f}(\omega), \label{7.18}%
\end{equation}
is a smooth function of $\omega$. Then%
\begin{align}
\mathsf{P}_{n}\cdot%
{\bm{F}}%
(t)  &  =\int_{0}^{t}ds\exp[-i\lambda_{n}(t-s)]f(s)\mathsf{P}_{n}\cdot%
{\bm{G}}%
_{0},\nonumber\\
\mathsf{P}_{ac}\cdot%
{\bm{F}}%
(t)  &  =\int_{0}^{t}ds\sum_{\alpha}\int d\lambda\exp[-i\lambda
(t-s)]f(s)|\mathbf{u}_{\lambda\alpha}\rangle\langle\mathbf{u}_{\lambda\alpha}|%
{\bm{G}}%
_{0}\rangle,\nonumber\\
\langle\mathbf{u}_{\lambda\alpha}|\mathsf{P}_{ac}\cdot%
{\bm{F}}%
(t)\rangle &  =\int_{0}^{t}ds\exp[-i\lambda(t-s)]f(s)\langle\mathbf{u}%
_{\lambda\alpha}|%
{\bm{G}}%
_{0}\rangle, \label{7.19}%
\end{align}
and%
\begin{align}
\mathsf{P}_{n}\cdot%
{\bm{F}}%
(t)  &  =\int_{0}^{t}ds\exp[-i\lambda_{n}(t-s)]\mathsf{P}_{n}\cdot\int
d\omega\exp[-i\omega s]%
{\tilde{\bm{G}}}%
(\omega)=\exp[-i\lambda_{n}t]\int d\omega\dfrac{\exp[i(\lambda_{n}%
-\omega)t]-1}{i(\lambda_{n}-\omega)}\tilde{f}(\omega)\mathsf{P}_{n}\cdot%
{\bm{G}}%
_{0}\nonumber\\
&  =\exp[-i\lambda_{n}t]\int dv\dfrac{1-\exp[-iv]}{-iv}\tilde{f}(\lambda
_{n}+\dfrac{v}{t})\mathsf{P}_{n}\cdot%
{\bm{G}}%
_{0}\overset{t\rightarrow\infty}{\sim}\dfrac{\pi}{2}\exp[-i\lambda_{n}%
t]\tilde{f}(\lambda_{n})\mathsf{P}_{n}\cdot%
{\bm{G}}%
_{0}, \label{7.20}%
\end{align}
whereas%
\begin{equation}
\langle\mathbf{u}_{\lambda\alpha}|\mathsf{P}_{ac}\cdot%
{\bm{F}}%
(t)\rangle\overset{t\rightarrow\infty}{\sim}\dfrac{\pi}{2}\exp[-i\lambda
t]\tilde{f}(\lambda)\langle\mathbf{u}_{\lambda\alpha}|%
{\bm{G}}%
_{0}\rangle,\;\parallel\mathsf{P}_{ac}\cdot%
{\bm{F}}%
(t)\parallel^{2}\overset{t\rightarrow\infty}{\rightarrow}(\dfrac{\pi}{2}%
)^{2}\sum_{\alpha}\int d\lambda|\tilde{f}(\lambda)|^{2}|\langle\mathbf{u}%
_{\lambda\alpha}|%
{\bm{G}}%
_{0}\rangle|^{2}. \label{7.21}%
\end{equation}
We see that for large times $\mathsf{P}_{n}\cdot%
{\bm{F}}%
(t)$ oscillates at the frequency $\omega_{n}$, a familiar situation for
musical instruments excited by a pulse (playing a piano, ringing a bell). We
have seen that in the dispersive half space case $\mathsf{K}$ has eigenvalues
$\pm\hat{\omega}$. Hence $%
{\bm{E}}%
(\mathbf{x},t)$ has contributions that oscillate at these frequencies. Whether
or not other contributions eventually die out depends on the nature of
$\tilde{f}(\lambda)$ and $\langle\mathbf{u}_{\lambda\alpha}|%
{\bm{G}}%
_{0}\rangle$. If the two have disjoint supports in $\lambda$, this will indeed
be the case. However, the continuous spectrum of $\mathsf{K}$ typically covers
the whole real axis.

Actual sources always have a finite bandwidth (although it can be quite small
as for single mode laser sources). In practice monochromatic sources (which
are usually simplified to a point source) are often considered. Thus if, for
instance, $f(t)=f_{0}\sin\omega_{0}t$, then $\tilde{f}(\lambda)=(2i)^{-1}%
f_{0}\{\delta(\lambda-\omega_{0})-\delta(\omega+\omega_{0})\}$. Now
$\mathsf{P}_{n}\cdot%
{\bm{F}}%
(t)$ vanishes if $\lambda_{n}\neq\pm\omega_{0}$ and diverges for $\lambda
_{n}=\pm\omega_{0}$. Also%
\begin{equation}
\langle\mathbf{u}_{\lambda\alpha}|\mathsf{P}_{ac}\cdot%
{\bm{F}}%
(t)\rangle=\dfrac{f_{0}}{2i}\exp[-i\lambda t]\{\dfrac{\exp[i(\lambda
-\omega_{0})t]-1}{i(\lambda-\omega_{0})}-\dfrac{\exp[i(\lambda+\omega
_{0})t]-1}{i(\lambda+\omega_{0})}\}\langle\mathbf{u}_{\lambda\alpha}|%
{\bm{G}}%
_{0}\rangle, \label{7.22}%
\end{equation}
and $\mathsf{P}_{ac}\cdot%
{\bm{F}}%
(t)$ diverges unless $\langle\mathbf{u}_{\lambda,\alpha}|%
{\bm{G}}%
_{0}\rangle$ vanishes in a neighborhood of $\pm\omega_{0}$. Such divergent
behavior is typical for undamped systems driven by a harmonic force. The
external source situation with a monochromatic point source is sometimes used
as the starting point for the calculation of Green's functions. Then, with $%
{\bm{G}}%
(\mathbf{x},t)=%
{\bm{G}}%
_{0}\exp[-i\omega_{0}t]\delta(\mathbf{x-x}_{0})$, taking Fourier transforms,%
\begin{equation}
\lbrack\omega-\mathsf{K}]\cdot%
{\tilde{\bm{F}}}%
(\omega)=i\delta(\omega-\omega_{0})\delta(\mathbf{x-x}_{0})%
{\bm{G}}%
_{0}. \label{7.23}%
\end{equation}
However, $\omega-\mathsf{K}$ does not have an inverse, $\omega$ being in the
spectrum of $\mathsf{K}$. This problem is avoided by using Laplace transforms,
in which case%
\begin{equation}%
{\hat{\bm{F}}}%
(z)=i[z-\mathsf{K}]^{-1}\cdot%
{\hat{\bm{G}}}%
(z),\;\operatorname{Im}z>0. \label{7.24}%
\end{equation}
Returning to the half space case, we conclude that for large times $%
{\bm{E}}%
(\mathbf{x},t)$ has oscillating contributions at the frequencies $\pm
\hat{\omega}$ but that there is also a additional contribution associated with
the continuous spectrum of $\mathsf{K}$.

\subsection{Final remarks}

The philosophy of our approach is to study special properties that occur for
specific frequencies, the NIM case being the primary example. But, as already
noted, in the half space case with a single scalar Lorentz contribution,
transmission tends to $0$ if $\omega\rightarrow\omega_{0}$. For this to
happen, $\varepsilon(z)$ and $\mu(z)$ need not be identical, as can be seen
from the expression for the Green's function. This feature remains valid if
there are more Lorentz contributions present. It suggests the experimental
study of reflection properties as a function of frequency. Dips will occur if
a NIM situation can happen (absorption will prevent obtaining an exact zero)
and maxima at the frequencies $\omega_{0n}$ in Eq. (7.2). In the idealized
single Lorentz case $\omega_{0}$ and $\hat{\omega}$ can both be obtained in
this way.

Although we concentrated on the situation where the electric and magnetic
susceptibilities were given by the same single dispersive Lorentz form, the
situation where they consist of different sets of Lorentz contributions can
also give rise to NIM situations, the fundamental requirement being the
existence of some frequencies $\hat{\omega}$ for which $\varepsilon
(\hat{\omega})=\mu(\hat{\omega})=-1$.

An important question is in how far such systems can be realized. On a
microscopic level it seems not to be possible to obtain this property (for
some further considerations on the susceptibilities of magnetodielectric
systems, see \cite{Raabe}). On a larger scale, small structures, involving
split rings and other configurations, lead to a more favorable situation. But
if the electric and magnetic modes are coupled, complications arise (von
Neumann's non-crossing rule). Another point is that, due to the employed
manufacturing methods, fabricated materials are not isotropic. But this may
change in the future. There is an extensive experimental literature concerning
the fabrication of such devices \cite{Experimental}.\textbf{ }A further
complication is often the occurrence of losses. The latter can spoil the
delicate effects essential for "perfect lenses ". In principle such lenses can
consist of a single NIM slab. In the present work we did not consider this
case, a simple form of a layered system. However, in a quite recent
publication, Collin \cite{Collin} made a precise analysis of this specific
case. He took into account field contributions not considered by Pendry
\cite{Pendry} and the effect of fields switched on for only a finite time
interval and found that taking these into account spoils the perfect lens
behavior. We also encountered such contributions in the half space case, the
background part in addition to the pole terms in the Green's functions.

Losses will blur the NIM behavior originating from pole contributions in a
dispersive case, since the poles now acquire an imaginary part. This raises
the question if adding some gain can improve the situation. Typically losses
arise from a coupling of the electromagnetic field to material modes that have
continuous spectrum. As an example, if the material consists of a single atom,
this coupling is the one to atomic continuum states (ionization). In
macroscopic media, among other possibilities, couplings to phonon modes and
the occurrence of F\"{o}rster processes can cause absorption. In such
situations electromagnetic energy is converted to material modes where the
energy leaks away in space. It will be clear that compensating all losses by
means of adding gain (this would convert the system to a dispersive,
non-absorptive one) will not be possible. But compensating loss at a few
specific frequencies, for instance by pumping the system to create level
inversions in the material subsystem, may be feasible. Here the gain must be
controlled precisely in order to maintain passivity and avoid undesirable
instabilities. This situation was analyzed by Stockman \cite{Existence},
\cite{Comments}, who applied a causality argument to the square of the
refraction function $n(z)^{2}=\varepsilon(z)\mu(z)$. He concluded that
negative refraction cannot be accomplished by adding gain.

Finally we mention an approach based on space-time transformations leading to
a simpler set of field equations but in a curved space-time frame
\cite{Transformations}. In particular the detailed work by Leonhardt and
Philbin, using methods borrowed from general relativity, should be mentioned.
So far this approach is restricted to frequency-independent $\varepsilon$ and
$\mu$. Taking the auxiliary field formalism as a starting point it may be
possible to extend it to the general frequency-dependent case.

\begin{acknowledgments}
The work of B. Gralak was partly supported by the project FANI
(ANR-07-NANO-038-03) of the program PNANO funded by the Agence Nationale de la
Recherche A. Tip was supported by FOM with financial support by NWO.
\end{acknowledgments}

\appendix

\section{Selfadjointness of \textsf{K}}

The idea is to split \textsf{K} into a zero order part \textsf{K}$_{0}$ and a
perturbation \textsf{K}$_{1}$,%
\begin{equation}
\mathsf{K}=\mathsf{K}_{0}+\mathsf{K}_{1}, \label{A1}%
\end{equation}
where%
\begin{equation}
\mathsf{K}_{0}=\left(
\begin{array}
[c]{cccccc}%
0 & 0 & 0 &
\bm{\epsilon}%
\cdot\mathbf{p} & 0 & 0\\
0 & 0 & 0 & 0 & 0 & i\lambda\\
0 & 0 & 0 & 0 & -i\lambda & 0\\
-%
\bm{\epsilon}%
\cdot\mathbf{p} & 0 & 0 & 0 & 0 & 0\\
0 & 0 & i\lambda & 0 & 0 & 0\\
0 & -i\lambda & 0 & 0 & 0 & 0
\end{array}
\right)  ,\;\mathsf{K}_{1}=\left(
\begin{array}
[c]{cccccc}%
0 & 0 & 0 & 0 & 0 & -i\langle%
\bm{\nu}%
_{e}|\\
0 & 0 & 0 & 0 & 0 & 0\\
0 & 0 & 0 & i & 0 & 0\\
0 & 0 & -i\langle%
\bm{\nu}%
_{m}| & 0 & 0 & 0\\
0 & 0 & 0 & 0 & 0 & 0\\
i & 0 & 0 & 0 & 0 & 0
\end{array}
\right)  . \label{A2}%
\end{equation}
\textit{Proposition:} Assume that $\parallel%
{\bm{\chi}}%
_{e,m}^{\prime}(\mathbf{x},0)\parallel_{\infty}=\sup_{\mathbf{x}}|%
{\bm{\chi}}%
_{e,m}^{\prime}(\mathbf{x},0)|\leqslant c<\infty$. Then \textsf{K}$_{1}$ is
bounded so \textsf{K} is selfadjoint with domain $\mathcal{D}(\mathsf{K}_{0}%
)$.\newline\textit{Proof:} For notational simplicity we give the proof for
scalar susceptibilities and the absorptive case (so $\nu_{e,m}(\mathbf{x}%
,\lambda)^{1/2}$ are properly defined, the dispersive situation must be
handled slightly differently). Let $\mathbf{f}\in\mathcal{D}(\mathsf{K}_{0})$.
Then%
\begin{equation}
\mathbf{g}=\left(
\begin{array}
[c]{c}%
\mathbf{g}_{1}\\
\mathbf{g}_{2}\\
\mathbf{g}_{3}\\
\mathbf{g}_{4}\\
\mathbf{g}_{5}\\
\mathbf{g}_{6}%
\end{array}
\right)  =\mathsf{K}_{1}\cdot\mathbf{f}=\left(
\begin{array}
[c]{c}%
-i\int d\lambda\nu_{e}(\mathbf{x},\lambda)\mathbf{f}_{6}(\mathbf{x},\lambda)\\
0\\
i\mathbf{f}_{4}(\mathbf{x})\\
-i\int d\lambda\nu_{m}(\mathbf{x},\lambda)\mathbf{f}_{3}(\mathbf{x},\lambda)\\
0\\
i\mathbf{f}_{1}(\mathbf{x})
\end{array}
\right)  . \label{A3}%
\end{equation}
Now%
\begin{align}
&  \parallel\mathbf{g}_{1}\parallel_{1}^{2}=\int d\mathbf{x}\int d\lambda
\nu_{e}(\mathbf{x},\lambda)\mathbf{f}_{6}(\mathbf{x},\lambda)\cdot\int d\mu
\nu_{e}(\mathbf{x},\mu)\overline{\mathbf{f}_{6}(\mathbf{x},\mu)}\nonumber\\
&  =\int d\mathbf{x}\int d\lambda\nu_{e}(\mathbf{x},\lambda)^{1/2}\int d\mu
\nu_{e}(\mathbf{x},\mu)^{1/2}\nu_{e}(\mathbf{x},\lambda)^{1/2}\mathbf{f}%
_{6}(\mathbf{x},\lambda)\cdot\nu_{e}(\mathbf{x},\mu)^{1/2}\overline
{\mathbf{f}_{6}(\mathbf{x},\mu)}\nonumber\\
&  \leqslant\int d\mathbf{x[}\int d\lambda\nu_{e}(\mathbf{x},\lambda
)]^{1/2}[\int d\mu\nu_{e}(\mathbf{x},\mu)]^{1/2}[\int d\lambda\nu
_{e}(\mathbf{x},\lambda)|\mathbf{f}_{6}(\mathbf{x},\lambda)|^{2}]^{1/2}[\int
d\mu\nu_{e}(\mathbf{x},\mu)|\mathbf{f}_{6}(\mathbf{x},\mu)|^{2}]^{1/2}%
\nonumber\\
&  =\int d\mathbf{x}\int d\lambda\nu_{e}(\mathbf{x},\lambda)\int d\lambda
\nu_{e}(\mathbf{x},\lambda)|\mathbf{f}_{6}(\mathbf{x},\lambda)|^{2}=\int
d\mathbf{x}\chi^{\prime}(\mathbf{x},0)\int d\lambda\nu_{e}(\mathbf{x}%
,\lambda)|\mathbf{f}_{6}(\mathbf{x},\lambda)|^{2}\leqslant\parallel
\chi^{\prime}(0)\parallel_{\infty}\parallel\mathbf{f}_{6}\parallel_{6}%
^{2}\leqslant d\parallel\mathbf{f}_{6}\parallel_{6}^{2}. \label{A4}%
\end{align}
Also%
\begin{equation}
\parallel\mathbf{g}_{3}\parallel_{3}^{2}=\int d\mathbf{x}\int d\lambda\nu
_{m}(\mathbf{x},\lambda)|\mathbf{f}_{4}(\mathbf{x})|^{2}\leqslant
d\parallel\mathbf{f}_{4}\parallel_{4}^{2}, \label{A5}%
\end{equation}
and similar for the other components. Thus%
\begin{equation}
\parallel\mathsf{K}_{1}\cdot\mathbf{f}\parallel\leqslant\sqrt{d}%
\parallel\mathbf{f}\parallel, \label{A6}%
\end{equation}
so $\mathsf{K}_{1}$ is a bounded selfadjoint operator and hence $\mathsf{K}$
is selfadjoint with domain $\mathcal{D}(\mathsf{K}_{0})$.$\blacksquare
$\medskip\newline\textit{Remark:} Note that the proof does not require
$\nu_{e,m}(\mathbf{x},\lambda)$ to be non-negative or even real. However, if
this is not the case the inner product on $\mathcal{K}$ is altered and the
norm no longer non-negative.

\section{Projections of $\mathsf{R}(z)$}

We assume the susceptibilities to be scalar and consider $\mathsf{P}%
_{1}[z-\mathsf{K}]^{-1}\mathsf{P}_{1}=\mathsf{P}_{1}\mathsf{R}(z)\mathsf{P}%
_{1}$, where \textsf{P}$_{j}$ projects upon the $j^{th}$ component of
$\mathbf{f}\in\mathcal{K}$, \textsf{P}$_{j}\mathbf{f}=\mathbf{f}_{j}$.We have%
\begin{align}
\lbrack z-\mathsf{K}]^{-1}  &  =[z+\mathsf{K}][z^{2}-\mathsf{K}^{2}%
]^{-1}=[z+\mathsf{K}]\left(
\begin{array}
[c]{ll}%
\lbrack z^{2}-\mathsf{K}_{em}\cdot\mathsf{K}_{me}]^{-1} & 0\\
0 & [z^{2}-\mathsf{K}_{me}\cdot\mathsf{K}_{em}]^{-1}%
\end{array}
\right) \nonumber\\
&  =[z+\mathsf{K}]\left(
\begin{array}
[c]{ll}%
\lbrack z^{2}-\mathsf{H}_{e}]^{-1} & 0\\
0 & [z^{2}-\mathsf{H}_{m}]^{-1}%
\end{array}
\right)  ,\nonumber\\
\mathsf{H}_{e}  &  =\left(
\begin{array}
[c]{lll}%
\mathsf{h}_{0}+\chi_{e}^{\prime}(\mathbf{x},0) & -\langle\nu_{e}|\lambda & -i%
\bm{\epsilon}%
\cdot\mathbf{p}\langle\nu_{m}|\\
-\lambda & \lambda^{2} & 0\\
-i%
\bm{\epsilon}%
\cdot\mathbf{p} & 0 & \lambda^{2}+|0\rangle\langle\nu_{m}|
\end{array}
\right)  ,\;\mathsf{H}_{m}=\left(
\begin{array}
[c]{lll}%
\mathsf{h}_{0}+\chi_{m}^{\prime}(\mathbf{x},0) & -\langle\nu_{m}|\lambda & i%
\bm{\epsilon}%
\cdot\mathbf{p}\langle\nu_{e}|\\
-\lambda & \lambda^{2} & 0\\
i%
\bm{\epsilon}%
\cdot\mathbf{p} & 0 & \lambda^{2}+|0\rangle\langle\nu_{e}|
\end{array}
\right)  , \label{B1}%
\end{align}
so%
\begin{equation}
\mathsf{P}_{1}[z-\mathsf{K}]^{-1}\mathsf{P}_{1}=z\mathsf{P}_{1}[z^{2}%
-\mathsf{H}_{e}]^{-1}\mathsf{P}_{1}. \label{B2}%
\end{equation}
According to the Feshbach projection formula with $A$ an operator and $P=1-Q$
a projector,%
\begin{align}
A^{-1}  &  =[QAQ]^{-1}Q+\{P-[QAQ]^{-1}QAP\}\mathcal{G}_{P}\{P-PAQ[QAQ]^{-1}%
\},\nonumber\\
PA^{-1}P  &  =\mathcal{G}_{P}P,\;\mathcal{G}_{P}=[PAP-PAQ(QAQ)^{-1}%
QAP]^{-1},\nonumber\\
QA^{-1}Q  &  =\mathcal{G}_{Q}Q,\;\mathcal{G}_{Q}=[QAQ-QAP(PAP)^{-1}PAQ]^{-1}.
\label{B3}%
\end{align}
In our case $P=\mathsf{P}_{1}=1-\mathsf{Q}_{1}=1-\mathsf{P}_{2}-\mathsf{P}%
_{3}$, $A=z^{2}-\mathsf{H}_{e}$, and we obtain%
\begin{equation}
\mathsf{P}_{1}[z^{2}-\mathsf{H}_{e}]^{-1}\mathsf{P}_{1}=[z^{2}-\mathsf{h}%
_{0}-\chi_{e}^{\prime}(\mathbf{x},0)-\mathsf{P}_{1}\mathsf{H}_{e}%
\mathsf{Q}_{1}[z^{2}-\mathsf{Q}_{1}\mathsf{H}_{e}\mathsf{Q}_{1}]^{-1}%
\mathsf{Q}_{1}\mathsf{H}_{e}\mathsf{P}_{1}]^{-1}\mathsf{P}_{1}. \label{B4}%
\end{equation}
Here%
\begin{align}
&  \mathsf{P}_{1}\mathsf{H}_{e}\mathsf{Q}_{1}[z^{2}-\mathsf{Q}_{1}%
\mathsf{H}_{e}\mathsf{Q}_{1}]^{-1}\mathsf{Q}_{1}\mathsf{H}_{e}\mathsf{P}%
_{1}=\mathsf{P}_{1}\mathsf{H}_{e}\mathsf{P}_{2}[z^{2}-\mathsf{P}_{2}%
\mathsf{H}_{e}\mathsf{P}_{2}]^{-1}\mathsf{P}_{2}\mathsf{H}_{e}\mathsf{P}%
_{1}+\mathsf{P}_{1}\mathsf{H}_{e}\mathsf{P}_{3}[z^{2}-\mathsf{P}_{3}%
\mathsf{H}_{e}\mathsf{P}_{3}]^{-1}\mathsf{P}_{3}\mathsf{H}_{e}\mathsf{P}%
_{1}\nonumber\\
&  =\langle\nu_{e}|\lambda^{2}[z^{2}-\lambda^{2}]^{-1}|0\rangle-%
\bm{\epsilon}%
\cdot\mathbf{p}\langle\nu_{m}|[z^{2}-\lambda^{2}-|0\rangle\langle\nu
_{m}|]^{-1}|0\rangle\cdot%
\bm{\epsilon}%
\cdot\mathbf{p}, \label{B5}%
\end{align}
and%
\begin{align}
&  \langle\nu_{e}|\lambda^{2}[z^{2}-\lambda^{2}]^{-1}|0\rangle=-\chi
_{e}^{\prime}(\mathbf{x},0)-z^{2}\hat{\chi}_{e}(\mathbf{x},z),\nonumber\\
&  \langle\nu_{m}|[z^{2}-\lambda^{2}-|0\rangle\langle\nu_{m}|]^{-1}%
|0\rangle\nonumber\\
&  =-1+[1-\langle\nu_{m}|[z^{2}-\lambda^{2}]^{-1}|0\rangle]^{-1}\nonumber\\
&  =-1+\mu(\mathbf{x},z)^{-1}, \label{B6}%
\end{align}
leading to%
\begin{align}
&  \mathsf{P}_{1}[z^{2}-\mathsf{H}_{e}]^{-1}\mathsf{P}_{1}\nonumber\\
&  =[z^{2}\varepsilon(\mathbf{x},z)+(%
\bm{\epsilon}%
\cdot\mathbf{p})\mu(\mathbf{x},z)^{-1}\cdot(%
\bm{\epsilon}%
\cdot\mathbf{p})\mathbf{]}^{-1}\mathsf{P}_{1}\nonumber\\
&  =\mathsf{R}_{e}(z)\mathsf{P}_{1}, \label{B7}%
\end{align}
so%
\begin{equation}
\mathsf{P}_{1}[z-\mathsf{K}]^{-1}\mathsf{P}_{1}=z\mathsf{R}^{e}(z)\mathsf{P}%
_{1}. \label{B8}%
\end{equation}
Similarly%
\begin{equation}
\mathsf{P}_{4}[z-\mathsf{K}]^{-1}\mathsf{P}_{4}=z\mathsf{R}^{m}(z)\mathsf{P}%
_{4}. \label{B9}%
\end{equation}

\section{Decomposition of the Green's functions for the layered case}

\noindent We express $\mathsf{R}_{%
{\bm{\kappa}}%
}^{e,m}(z)$ in terms of the inverses of scalar operators. Using the Feshbach
formula, Eq. (B3), with $A=$ $\mathsf{L}_{%
{\bm{\kappa}}%
}^{e,m}(z)$ and \ \
\begin{align}
\mathsf{P}_{s}  &  =\mathbf{e}_{3}\times\mathbf{e}_{%
{\bm{\kappa}}%
}\mathbf{e}_{3}\times\mathbf{e}_{%
{\bm{\kappa}}%
},\nonumber\\
\mathsf{Q}_{s}  &  =\mathsf{U}-\mathsf{P}_{s}=\mathbf{e}_{%
{\bm{\kappa}}%
}\mathbf{e}_{%
{\bm{\kappa}}%
}+\mathbf{e}_{3}\mathbf{e}_{3}, \label{C1}%
\end{align}
we find, noting that%
\begin{align}
&  (\mathbf{e}_{3}\times\mathbf{e}_{%
{\bm{\kappa}}%
})\cdot(%
\bm{\epsilon}%
\cdot\mathbf{p)}=\kappa\mathbf{e}_{3}-p\mathbf{e}_{%
{\bm{\kappa}}%
},\nonumber\\
&  (%
\bm{\epsilon}%
\cdot\mathbf{p)\cdot}(\mathbf{e}_{3}\times\mathbf{e}_{%
{\bm{\kappa}}%
})=-\kappa\mathbf{e}_{3}+p\mathbf{e}_{%
{\bm{\kappa}}%
}\nonumber\\
&
\bm{\epsilon}%
\cdot\mathbf{p)}\cdot\mathsf{Q}_{s}\nonumber\\
&  =(%
\bm{\epsilon}%
\cdot\mathbf{p)}\cdot(\mathbf{e}_{%
{\bm{\kappa}}%
}\mathbf{e}_{%
{\bm{\kappa}}%
}+\mathbf{e}_{3}\mathbf{e}_{3})\nonumber\\
&  =(\mathbf{e}_{3}\times\mathbf{e}_{%
{\bm{\kappa}}%
})(\kappa\mathbf{e}_{3}-p\mathbf{e}_{%
{\bm{\kappa}}%
}), \label{C2}%
\end{align}
etc., that (we skip the subscript $%
{\bm{\kappa}}%
$ for brevity) \ \
\begin{align}
&  \mathsf{P}_{s}\cdot\mathsf{L}^{e}\cdot\mathsf{P}_{s}=[\frac{\zeta^{2}}{\mu
}-p\frac{1}{\mu}p]\mathsf{P}_{s},\nonumber\\
&  \mathsf{P}_{s}\cdot\mathsf{L}^{e}\cdot\mathsf{Q}_{s}=0,\nonumber\\
&  \mathsf{Q}_{s}\cdot\mathsf{L}^{e}(z)\cdot\mathsf{Q}_{s}=\mathsf{L}_{p}%
^{e}\nonumber\\
&  =[z^{2}\varepsilon(x,z)\mathsf{U}_{Q}-(p\mathbf{e}_{%
{\bm{\kappa}}%
}-\kappa\mathbf{e}_{3})\frac{1}{\mu(x,z)}(p\mathbf{e}_{%
{\bm{\kappa}}%
}-\kappa\mathbf{e}_{3})]\cdot\mathsf{Q}_{s} \label{C3}%
\end{align}
and similar for $\mathsf{L}_{%
{\bm{\kappa}}%
}^{m}$ (interchange $\varepsilon$ and $\mu$), where%
\begin{equation}
\zeta(x,\kappa,z)^{2}=z^{2}\varepsilon(x,z)\mu(x,z)-\kappa^{2}. \label{C4}%
\end{equation}
Thus%
\begin{equation}
\mathsf{R}^{e,m}(z)=\mathsf{R}_{s}^{e,m}(z)+\mathsf{R}_{p}^{e,m}(z),
\label{C5}%
\end{equation}
where, with%
\begin{equation}
R_{s}^{e}(z)=[z^{2}\varepsilon-\frac{\kappa^{2}}{\mu}-p\frac{1}{\mu}%
p]^{-1},\;R_{s}^{m}(z)=[\frac{\zeta^{2}}{\varepsilon}-p\frac{1}{\varepsilon
}p]^{-1}, \label{C6}%
\end{equation}%
\begin{align}
\mathsf{R}_{s}^{e}(z)  &  =R_{s}^{e}(z)\mathsf{P}_{s},\;\mathsf{R}_{s}%
^{m}(z)=R_{s}^{m}(z)\mathsf{P}_{s},\nonumber\\
\mathsf{R}_{p}^{e}(z)  &  =[z^{2}\varepsilon\mathsf{U}_{Q}-(p\mathbf{e}_{%
{\bm{\kappa}}%
}-\kappa\mathbf{e}_{3})\frac{1}{\mu}(p\mathbf{e}_{%
{\bm{\kappa}}%
}-\kappa\mathbf{e}_{3})]^{-1}\mathsf{Q}_{s},\nonumber\\
\mathsf{R}_{p}^{m}(z)  &  =[z^{2}\mu\mathsf{U}_{Q}-(p\mathbf{e}_{%
{\bm{\kappa}}%
}-\kappa\mathbf{e}_{3})\dfrac{1}{\varepsilon}(p\mathbf{e}_{%
{\bm{\kappa}}%
}-\kappa\mathbf{e}_{3})]^{-1}\mathsf{Q}_{s}. \label{C7}%
\end{align}
Now let%
\begin{equation}
\mathsf{R}_{p}^{e}(z)=A\mathbf{e}_{%
{\bm{\kappa}}%
}\mathbf{e}_{%
{\bm{\kappa}}%
}+B\mathbf{e}_{%
{\bm{\kappa}}%
}\mathbf{e}_{3}+C\mathbf{e}_{3}\mathbf{e}_{%
{\bm{\kappa}}%
}+D\mathbf{e}_{3}\mathbf{e}_{3}. \label{C8}%
\end{equation}
Since%
\begin{equation}
\mathsf{L}_{p}^{e}\cdot\mathsf{R}_{p}^{e}=\mathsf{Q}_{s}=\mathbf{e}_{%
{\bm{\kappa}}%
}\mathbf{e}_{%
{\bm{\kappa}}%
}+\mathbf{e}_{3}\mathbf{e}_{3}, \label{C9}%
\end{equation}
we find by comparing coefficients that%
\begin{align}
(z^{2}\varepsilon-p\frac{1}{\mu}p)A+p\frac{\kappa}{\mu}C  &  =1,\nonumber\\
\frac{\zeta^{2}}{\mu}D+\frac{\kappa}{\mu}pB  &  =1,\nonumber\\
\frac{\kappa}{\mu}pA+\frac{\zeta^{2}}{\mu}C  &  =0,\nonumber\\
(z^{2}\varepsilon-p\frac{1}{\mu}p)B+p\frac{\kappa}{\mu}D  &  =0. \label{C10}%
\end{align}
Let%
\begin{equation}
R_{p}^{e}(z)=[z^{2}\varepsilon-p\frac{z^{2}\varepsilon}{\zeta^{2}}%
p]^{-1},\;R_{p}^{m}(z)=[z^{2}\mu-p\frac{z^{2}\mu}{\zeta^{2}}p]^{-1}.
\label{C11}%
\end{equation}
Then%
\begin{align}
A  &  =R_{p}^{e}(z),\;B=-R_{p}^{e}(z)p\frac{\kappa}{\zeta^{2}},\;C=-\frac
{\kappa}{\zeta^{2}}pR_{p}^{e}(z),\nonumber\\
D  &  =\frac{\mu}{\zeta^{2}}+\frac{\kappa}{\zeta^{2}}p[z^{2}\varepsilon
-p\frac{z^{2}\varepsilon}{\zeta^{2}}p]^{-1}p\frac{\kappa}{\zeta^{2}},
\label{C12}%
\end{align}
with similar results for the magnetic case. Hence%
\begin{align}
&  \mathsf{R}^{e}(z)=R_{s}^{e}(z)\mathbf{e}_{3}\times\mathbf{e}_{%
{\bm{\kappa}}%
}\mathbf{e}_{3}\times\mathbf{e}_{%
{\bm{\kappa}}%
}\nonumber\\
&  +(\mathbf{e}_{%
{\bm{\kappa}}%
}-\frac{\kappa}{\zeta^{2}}p\mathbf{e}_{3})R_{p}^{e}(z)(\mathbf{e}_{%
{\bm{\kappa}}%
}-p\frac{\kappa}{\zeta^{2}}\mathbf{e}_{3})+\frac{\mu}{\zeta^{2}}\mathbf{e}%
_{3}\mathbf{e}_{3},\nonumber\\
&  \mathsf{R}^{m}(z)=R_{s}^{m}(z)\mathbf{e}_{3}\times\mathbf{e}_{%
{\bm{\kappa}}%
}\mathbf{e}_{3}\times\mathbf{e}_{%
{\bm{\kappa}}%
}\nonumber\\
&  +(\mathbf{e}_{%
{\bm{\kappa}}%
}-\frac{\kappa}{\zeta^{2}}p\mathbf{e}_{3})R_{p}^{m}(z)(\mathbf{e}_{%
{\bm{\kappa}}%
}-p\frac{\kappa}{\zeta^{2}}\mathbf{e}_{3})+\frac{\varepsilon}{\zeta^{2}%
}\mathbf{e}_{3}\mathbf{e}_{3}. \label{C13}%
\end{align}
Note that%
\begin{equation}
R_{p}^{e}(z)=\frac{1}{p}\frac{\zeta^{2}}{z^{2}\varepsilon}R_{s}^{m}%
(z)p\frac{1}{\varepsilon}, \label{C14}%
\end{equation}
so $\mathsf{R}_{p}^{e}(z)$ can be expressed in terms of the scalar magnetic
$s$-polarized Green's function but the vectors in front and behind are quite
different. From the above expressions the corresponding Green's functions,
introduced in Sect. V, now follow.

\end{document}